\documentclass[12pt,preprint]{aastex}

\newcommand{\punkt}{\mathrm{ \hspace*{0.1cm}.}}
\newcommand{\mdot}{\ensuremath{\dot{M}}}
\newcommand{\teff}{\ensuremath{T_{\rm eff}}}
\newcommand{\rsun}{\ensuremath{\, {\rm R}_\odot}}
\newcommand{\lsun}{\ensuremath{\, {\rm L}_\odot}}
\newcommand{\msun}{\ensuremath{\, {\rm M}_\odot}}
\newcommand{\n}{\ensuremath{\mem{n}}}

\newcommand{\p}{\ensuremath{\mem{p}}}
\newcommand{\e}{\ensuremath{\mem{e}}}

\newcommand{\hedr}{\ensuremath{^{3}\mem{He}}}
\newcommand{\hevi}{\ensuremath{^{4}\mem{He}}}

\newcommand{\cdr}{\ensuremath{^{13}\mem{C}}}
\newcommand{\czw}{\ensuremath{^{12}\mem{C}}}

\newcommand{\nvi}{\ensuremath{^{14}\mem{N}}}
\newcommand{\nfu}{\ensuremath{^{15}\mem{N}}}

\newcommand{\ose}{\ensuremath{^{16}\mem{O}}}
\newcommand{\osi}{\ensuremath{^{17}\mem{O}}}
\newcommand{\oac}{\ensuremath{^{18}\mem{O}}}

\newcommand{\fac}{\ensuremath{^{18}\mem{F}}}

\newcommand{\nezwa}{\ensuremath{^{20}\mem{Ne}}}
\newcommand{\neei}{\ensuremath{^{21}\mem{Ne}}}
\newcommand{\nezw}{\ensuremath{^{22}\mem{Ne}}}

\newcommand{\nadr}{\ensuremath{^{23}\mem{Na}}}
\newcommand{\mgvi}{\ensuremath{^{24}\mem{Mg}}}
\newcommand{\mgfu}{\ensuremath{^{25}\mem{Mg}}}
\newcommand{\mgse}{\ensuremath{^{26}\mem{Mg}}}

\newcommand{\alse}{\ensuremath{^{26}\mem{Al}}}

\newcommand{\zrse}{\ensuremath{^{96}\mem{Zr}}}
\newcommand{\zrfu}{\ensuremath{^{95}\mem{Zr}}}
\newcommand{\zrvi}{\ensuremath{^{94}\mem{Zr}}}

\newcommand{\mem}[1]{\ensuremath{\mathrm{ #1}}}
\newcommand{\spr}{\mbox{$s$-process}}
\newcommand{\kelv}{\ensuremath{\,\mathrm K}}
\newcommand{\abb}[1]{Fig.\,\ref{#1}}
\newcommand{\kap}[1]{\S\,\ref{#1}}
\newcommand{\jahre}{\ensuremath{\, \mathrm{yr}}}
\newcommand{\glt}[1]{Eq.\,(\ref{#1})}
\newcommand{\glp}[1]{(Eq.\,\ref{#1})}
\newcommand{\tab}[1]{Table\,\ref{#1}}

\slugcomment{LA-UR: 04-4530}

\usepackage{natbib}

\shorttitle{EMP Intermediate mass stars}
\shortauthors{F. Herwig}

\received{}
\begin{document}

\title{Evolution and Yields of Extremely Metal Poor Intermediate Mass Stars}  
\author{Falk Herwig}
\affil{Los Alamos National Laboratory, Los Alamos, NM 87544} \email{fherwig@lanl.gov}
\affil{Department of Physics and Astronomy, University of Victoria,
  3800 Finnerty Rd, Victoria, BC, V8P 1A1}

\begin{abstract}
  Intermediate mass stellar evolution tracks from the main sequence to
  the tip of the AGB for five initial masses ($2$ to $6\msun$) and
  metallicity Z=0.0001 have been computed. The detailed 1D structure
  and evolution models include exponential overshooting, mass loss and
  a detailed nucleosynthesis network with updated nuclear reaction
  rates. The network includes a two-particle heavy neutron sink for
  approximating neutron density in the He-shell flash.  It is shown
  how the neutron-capture nucleosynthesis is important in models of
  very low metallicity for the formation of light neutron-heavy
  species, like sodium or the heavy neon and magnesium isotopes. The
  models have high resolution, as required for modeling the third
  dredge-up.  All sequences have been followed from the pre-main
  sequence to the end of the AGB when all envelope mass is lost.
  Detailed structural and chemical model properties as well as yields
  are presented. This set of stellar models is based on standard
  assumptions and updated input physics. It can be confronted with
  observations of extremely-metal poor stars and may be used to assess
  the role of AGB stars in the origin of abundance anomalies of
  some Globular Cluster members of correspondingly low metallicity.
\end{abstract}

\keywords{ stars: AGB 
--- stars: evolution
--- nuclear reactions, nucleosynthesis, abundances
--- stars: interiors}

\section{Introduction}
\label{sec:ov}
The chemical production of low- and intermediate mass stars during
their asymptotic giant branch phase of evolution is an important
contribution to the galactic chemical evolution. For example, half of
all elements heavier than mass number $A=90$ are made in low mass AGB
stars by the \spr. AGB stars are responsible for a significant
fraction of the carbon in the solar system abundance mix. AGB stars
are the favored candidate for the main primary nitrogen source at
lower metallicity \citep{pettini:02}. The discrepancy of observations
and galactic chemical evolution predictions based on massive star
yields is in particular evident for some neutron heavy isotopes of
certain elements \citep[like Mg or Na,][]{timmes:95}. This indicates
the role of very low metallicity AGB stars to the early galactic
chemical evolution. Equally important are the informations obtained
from an increasing number of spectroscopic observations of extremely
metal-poor (EMP) stars.

In the past yield predictions for a range of initial masses
and metallicities have been derived from so-called synthetic
AGB models \citep{renzini:81,marigo:96,hoek:97}. In these models the
evolution is described by constructing fitting formulas to certain
quantities of full stellar evolution models that solve the full set of
stellar structure and nuclear energy generation. The most recent
improvements include a full envelope integration for the hot bottom
burning phase of massive AGB stars \citep{marigo:98}. The synthetic
AGB model approach is very well justified under certain circumstances
and for specific purposes. The computation of AGB stellar models
requires high numerical resolution and significant computing time. In
the past large grids in mass and metallicity with several ten- to
onehundredthousand models for each sequence have not been feasible due
to their computationally demanding nature. This is  changing now
and as an example, \citet{karakas:02a} have presented a comprehensive
grid of AGB calculations to study differentially the dependence of DUP
on mass and metallicity. Still, full stellar evolution models in the
past were not able to reproduce, for example, the third dredge-up (DUP) in a way that agrees with observations.  New results from
nuclear physics may help to resolve this issue in the future
\citep{herwig:04e}. Synthetic models on the other hand can
parameterize the third DUP in some way, and calibrate its efficiency
with observed properties of AGB stars, for example using the C-star
luminosity function of the Magellanic Clouds. The synthetic model
predictions are therefore internally consistent with observations.  In
addition these models are useful because they condense the information
contained in the C-star LF, or in C-rich star and O-rich star number
counts in the Magellanic Clouds into more simple constraints, which
detailed structure and evolution models have to reproduce. For
example, the synthetic models by \citet{marigo:96} require that
efficient dredge-up must take place at core masses as low as
$0.58\msun$, a value which has even been revised downward by a more
recent update \citep{marigo:99}.

In order to model yields and the abundance evolution for EMP stars
full stellar models appear to be in particular important. We are just
beginning to identify the peculiarities of nucleosynthesis in EMP AGB
stars \citep[or at Z=0 should they have existed,][]{chieffi:01}.
 Synthetic models rely both on accurate
structure and evolution models to derive the fitting formulas, as well
as on a good calibration. Both are presently not well established at
metallicities of $Z<0.001$.  Models of massive AGB stars at
very low metallicity for example, show behaviors not seen in models
of larger metallicity, like H-burning during the DUP or DUP reaching
below the He-shell \citep{herwig:03c}. It is therefore not appropriate
to simply extrapolate synthetic models that have been calibrated at
moderate metal deficiency to very low metallicities. In order to
improve the situation more complete stellar evolution calculations are
needed. Here, the structure and chemical evolution for a set of five
tracks with masses from $2\msun$ to $6\msun$ at metallicity $Z=0.0001$
are presented. 

The application of these models may include comparison to observations
of EMP stars or tests of the possible role IMS (intermediate mass
stars) in the chemical evolution of low-metallicity Globular Clusters.
Although the grid is far from complete in mass and metallicity the
calculations may be useful for Galactic Chemical Evolution models.
Since the structural evolution is given as well, the models may be
used to investigate the s-process.

In Paper I \citep{herwig:03c} the evolution of massive AGB stars at
very low metallicity has been described in detail, in particular the
interplay of hot bottom burning and the \emph{hot} DUP. In another
previous paper we have specifically addressed the possible
implications of the more massive cases at this metallicity for the
star-to-star abundance variations observed in globular clusters
\citep{denissenkov:03a}.  Here the detailed thermodynamic, structural
and abundance evolution of a homogeneous set of intermediate mass
EMP stellar models is presented.
In \kap{sec:code} the code changes compared to the version used
in Paper I are described. The model results are given in
\kap{sec:results}. The conclusions
  are presented in \kap{sec:concl}. 

\section{Physical input, stellar evolution code and model calculations}
\label{sec:code}

The stellar evolution code and all model assumptions are the same as
in Paper I, with the following exceptions. 
\subsection{Mass loss}
Mass loss for AGB stars of very low metallicity is neither
observationally nor theoretically well constrained, and one has to
rely on extrapolation from larger metallicities and rather qualitative
and indirect considerations. For a mass loss formula both the
functional dependence of mass loss on stellar parameters as well as
the absolute calibration are needed.

Mass loss rates of Mira-like pulsating stars can be calculated by
modeling the dynamical structure and evolution of their atmospheres
\citep{bowen:88}. For solar-like metallicity \citet{bloecker:95a} has
extracted the luminosity dependence of the \citet{bowen:88} models and
finds 
\begin{equation}
\label{eq:blmdotl}
\log \mdot \propto 3.7 \log L \punkt
\end{equation}
 The physical input like the
opacities as well as the properties of dust formation of the dynamical
atmosphere models depend on the metallicities. However, no such models
are available for low or very low Z. Here I assume not only that
\glt{eq:blmdotl} can be applied at $Z=10^{-4}$, but also that it
approximately applies to changes in L due to changes in Z too. This
assumption may in fact not be entirely foolish. The
DUP is more efficient in EMP AGB stars \citep[][and
\kap{sec:results}]{herwig:03c} and the CNO elements collectively reach
close to the solar value abundances in the envelope. A comparison of
the Rosseland mean opacity for a 3\msun\ TP-AGB model at $Z=0.02$ and
$Z=10^{-4}$ yields a difference of typically less than $10\%$ for
$\log T < 4.6$.

If the dependence \glt{eq:blmdotl} is applied to luminosity variation
due to varying metallicity, then  the variation of
stellar parameters with metallicity introduces a significant variation
of the mass loss. In fact due to this effect  one obtains a larger mass loss at
lower Z because 
the luminosity is larger at lower Z. The average luminosity and
temperature of a 2\msun\ and a 5\msun\ for a range of metallicities is
shown in \abb{fig:hrd-z}. From these stellar structure and evolution
results one
determines approximately  
\begin{equation}
\label{eq:evolz}
\log L \propto -0.2 \log Z \punkt
\end{equation}
Together with \glt{eq:blmdotl} this implies
that $\mdot(Z=10^{-4})\sim \eta_\mem{1} \cdot \mdot(Z=0.02)$ with
$\eta_\mem{1}\sim 50$. 

\citet{vanloon:00} has attempted to derive
a mass loss-metallicity relation based on observations of Magellanic Cloud
giants, and suggests that 
\begin{equation}
\label{eq:vanloon}
\log \mdot \propto 0.3 \log Z \punkt
\end{equation}
This exponent has a very large uncertainty, mainly due to small-number
statistics. Due to its observational nature \glt{eq:vanloon} contains
both the possible dependence of \glt{eq:blmdotl} on metallicity and
the dependence of stellar parameters on metallicity \glp{eq:evolz}.
Tentatively extrapolating the  relation of \citet{vanloon:00} to
$Z=10^{-4}$ the absolute  mass loss rates at $Z=0.02$ and
$Z=10^{-4}$  are related by $\mdot(Z=10^{-4})\sim
\eta_\mem{2} \cdot \mdot(Z=0.02)$ with $\eta_\mem{2}\sim 0.2$. 

One can then obtain a calibration for the Bowen-Bloecker mass loss
rate \glp{eq:blmdotl} at $Z=10^{-4}$ by $\eta_\mem{B}=\eta_\mem{2}/\eta_\mem{1}$ 
which gives  $\eta_\mem{B}=0.004$ for the exponent $0.3$ of
\glt{eq:vanloon}. However, it has been mentioned that this exponent is
subject to large uncertainty. In particular, both result for M stars
and for C stars are compatible with $\log \mdot \propto -0.3 \log Z$
which implies $\eta_\mem{2}\sim 5$. Therefore the largest mass loss at
$Z=10^{-4}$ within this framework is  $\eta_\mem{B}=0.1$ and this is
the adopted value for the present study. In this way these calculations
give a lower limit to the envelope enrichment and the 
corresponding yields. A lower mass loss rate would lead to a longer
TP-AGB duration with more TPs and subsequent DUP events.

\subsection{Overshooting}
The treatment of convective boundaries, both at the bottom of the
envelope convection as well as at the bottom of the pulse-driven
convection zone (PDCZ) during the He-shell flash effects stellar
models in many ways, including the dredge-up predictions. The
consequences and implications of various treatments of mixing at these
convective boundaries has been explored in detail
\citep{herwig:97,mowlavi:99,herwig:99a,langer:99,lugaro:02a,herwig:02a,herwig:03c}.
Models which include an adjustable amount of depth- and time-dependent
amount of overshooting at all convective boundaries show efficient
dredge-up at low core-masses.

The same concept of exponential overshooting as in Paper I is applied.
For the TP-AGB models the overshoot efficiency at the bottom of the
PDCZ is set to $f_\mem{PDCZ}=0.008$ while the efficiency at the bottom
of the convective envelope is set to $f_\mem{CE}=0.016$ at all times.
Models with larger $f_\mem{PDCZ}$ have higher temperature for the
\nezw\ neutron source.  \citet{lugaro:02a} found that if
$f_\mem{PDCZ}$ is too large some s-process branchings produce
isotopic ratios that are irreconcilable with laboratory measurements
in pre-solar SiC grains. For example, the branching at \zrfu\ is
sensitive to the temperature at the bottom of the He-shell flash
convection zone. If $f_\mem{PDCZ}$ is too large the temperature is too
large and the predicted \zrse/\zrvi\ ratio is larger than measured in
the grains.  \citet{lugaro:02a} have concluded that
$f_\mem{PDCZ}=0.016$ may be too large. Unfortunately the neutron cross
section of the unstable isotope \zrfu\ is very uncertain. The current
range of estimates differ within a factor of four \citep[Kawano,
priv.\ com.,][]{bao:00,jorissen:01}, and a more conclusive analysis
has to await a better experimental determination of this cross
section. Other nuclear reaction rates as studied by \citet{herwig:04e}
may also be important. In any case the concept of overshooting at the
bottom of the PDCZ plays an important role in currently favored models
of H-deficient central stars of planetary nebulae \citep[{of spectral
  type PG1159 and
  [WC]-CSPN,}][]{koesterke:97b,dreizler:96,werner:05a}.

The rather small value for $f_\mem{CE}=0.016$ can not reproduce the
partial mixing needed for the \cdr\ neutron source in low mass TP-AGB
stars \citep{goriely:00,herwig:02a}. One can estimate the mass of the
s-process layer resulting from partial mixing that reproduces the
observed stellar s-process overabundances. The stellar model must
accomplish such an enrichment within a number of dredge-up events that can be constrained by observations. This number can be derived from
model comparison with the observed C-star luminosity function and the
observed ratio of C-stars to O-rich AGB stars \citep{marigo:99}. In
particular newer synthetic models including the effect of C/O ratio
dependent molecular opacities \citep{marigo:02} indicate that typical
low-mass AGB stars that produce the s-process elements may in fact
experience only five to at most ten thermal pulses and subsequent
dredge-up events during the C-rich phase.

In the context of evaluating the properties of rotating AGB star
models \citet{herwig:02a} estimated that the partial mixing zone that
hosts the s-process nucleosynthesis should have a mass
$M_\mem{P}>7\cdot10^{-5}\msun$, in agreement with more detailed
calculations. Exponential overshooting as applied here with an
efficiency $f_\mem{CE}=0.016$ leads to a partial mixing zone with a
mass $\approx 10^{-6}\msun$ only \citep{herwig:99a,herwig:02a}. In a
detailed analysis \citet{lugaro:02a} have assumed that
$f_\mem{CE}=0.128$ and obtained overall overabundance factors close to
what is observed.  There are indications
that $f_\mem{CE}$ may be even larger in the framework of
the \spr\ partial-mixing concept, at least for solar-metallicity
cases. 

However, in Paper I it was shown that in models of extremely low
metallicity even a very small amount of exponential overshooting leads
to \emph{hot} DUP with potentially very significant implications to
the overall evolution. In these stars the \czw\ rich intershell
material beneath the proton-rich matter is hotter than in cases with
higher metallicity. If protons are forced into the \czw\ rich core
during the dredge-up phase of the TP cycle they are burnt vigorously.
The additional H-burning luminosity can further drive convective
instability and cause a corrosive, convective H-shell that thus
penetrates much deeper into the core than without this additional
H-burning. This \emph{hot DUP} and the resulting corrosive burning
is in fact conceptually related to the conductive propagation of
nuclear flames considered in ONeMg cores in the context of SN Type Ia
models \citep{timmes:94}.

In Paper I calculations of a $5\msun$ dredge-up phase with
$f_\mem{CE}=0.03$ -- about twice as large as the value assumed here --
prompted ongoing dredge-up that would have terminated the AGB in
approximately $2000\jahre$ (based on some uncertain assumptions on
mass loss for such configurations). The situation for lower-mass
cases may be less dramatic, but still hot DUP may alter the
conditions for the formation of a partial mixing zone at extremely low
metallicity, as enhanced H-burning during this phase would result from
any efficient convection-driven extra-mixing process.

While it is possible that exponential extra-mixing as assumed in the
diffusive overshooting approach is responsible for the \spr\ partial
mixing zone at solar metallicity, the situation is rather unclear at
extremely low metallicity. The situation can be improved by
investigating how the structure (i.e.\ mainly the DUP depth) depends
on the interplay of extra-mixing and hot DUP, and how extra-mixing
itself is effected by nuclear burning in the overshooting layer.  In
these calculations a rather small efficiency of overshooting is
assumed. This way most of the complications caused by hot DUP are
avoided. However, the nuclear production predictions do not include
the uncertain effect of a more extended \cdr\ pocket.

\subsection{Nuclear network}
The updated nuclear reaction network
includes all relevant charged particle reactions, $\beta$-decays and
neutron-capture reactions. If available the NACRE reaction rates are used
(\tab{tab:nuc-netw}). The nuclear network solution is based on rates from 
linear interpolations in the 250 grid point logarithmic tables generated by
the WWW database tool \emph{NETGEN} \citep{jorissen:01}. For the
calculation of the energy generation  in the stellar structure solution
a smaller network containing the dominating CNO cycle, pp-chain, and
He-burning reactions has been used. The reaction rates  have been
calculated from the fitting formulas to the NACRE reactions
\citep{angulo:99}, which provide smooth T-derivatives. Potentially
this dual approach may introduce an inconsistency because a small
amount of energy released by trace elements may be ignored, and
because the fitting formulas may disagree with the tabulated values by
a few percent. However, it turns out that the fitting formula
(despite the published fitting errors) represent the tabulated values
very well for the most important H- and He-burning reactions in the
relevant temperature range. 

All important light n-capture reactions as well as the several
Fe-group species capturing neutrons are included. A light and a heavy
n-sink take care of the remaining n-capture species as in
\citet{herwig:02a}. The two-particle sink approximating the trans-iron
species allows a rough estimate on the number of neutrons captured per
iron seed particle ($n_\mem{cap}$) which is an important diagnostic
tool for the \spr. For the sink-cross sections the same value as in
\citet{herwig:02a} is taken.  This cross section depends on the heavy
elements abundance distribution, and therefore a cross section derived
for (near-)solar metallicity is not correct at very low metallicity.
However, a detailed study of the \spr\ at very low metallicity which
is required to estimate an appropriate sink-cross section is not
available at this time. In addition the exact value of the heavy
sink-cross section is less important at very low metallicity because
the role of heavy species for the determination of the neutron density
is smaller when large amounts of primary light n-capture species, like
\nezw, are present.
 
\section{Results}
\label{sec:results}
\subsection{Structural evolution}
The average structural properties of the model grid are summarized in
\tab{tab:cm-l}. It shows how the stellar parameters as well as
evolutionary times change systematically as a function of initial
mass. 

The surface temperature of the models as a function of TP-phase and
mass changes only little, and all TP-AGB models are in the range
$\teff=4000 - 4350\kelv$. All sequences show very efficient third
dredge-up. Therefore the core masses are on average almost constant
during the TP-AGB evolution. These full stellar evolution models
do not support the scenario of low-mass supernovae in the
early Galactic halo proposed by \citet{zijlstra:04}. An important
assumptions in this proposal is that the cores of IMS stars are able
to grow to the Chandrasekhar mass because the mass loss is very low at
low metallicity. The models presented here can not rule out this
proposal, but it is fair to say that they do at least raise some
serious doubts. Deep dredge-up events prevent the core from growing.
They also cause substantial pollution of the surface layers with C, N
and O, all of which are important sources of opacity in the outer
stellar layers. Thus the assumption that mass loss is in fact
negligible in these stars may not be correct. Finally, the hot DUP
(previous section and Paper I) is more effective at higher core mass.
Thus, from these calculations I have
more reasons to doubt rather than believe this idea. 

 The stellar luminosity and radius evolve
only little with TP number and during a TP cycle. Averaged values are
given in \tab{tab:cm-l}. These TP-AGB models with initial masses of $2
\dots 6\msun$ have core masses in the range $0.62 \dots 1.05\msun$.
The luminosities follow approximately an exponential core-mass
luminosity relation:
\begin{equation} \label{gl:mc-l}
\log L/\lsun = 1.7414 \times M_\mem{c}/\msun + 2.8799
\end{equation}
This is shown in \abb{fig:cm-l}, where in addition two data points for
TP-AGB stars (last TP cycle) of zero-metallicity by \citet{chieffi:01}
are shown. Our models show sharp luminosity and radius peaks during
the first deep dredge-up events (see Paper I for details). These peaks
may be important if the binary evolution and of EMP stars (in the
CH-star scenario) is considered. It may be that interaction or
enhanced tidal synchronization occurs preferentially during these
spikes close to the time of the He-shell flash.

The core mass is defined here as the mass within which the hydrogen
mass fraction  is $X_\mem{H}<0.37$. The core mass grows because
of H-shell burning. The recurrent third dredge-up events  decrease
the core mass. The result is a characteristic saw-tooth shaped curve
of the core-mass as a function of time. In \abb{fig:t-M} the core mass
evolution is shown for all sequences.  The DUP parameter is defined as
$\lambda=\Delta M_\mem{H}/\Delta M_\mem{DUP}$, where $\Delta
M_\mem{H}$ is the core mass growth due to H-burning during the
interpulse phase. In all sequences $\lambda \sim 1$ for the majority of
TPs. Models with larger core mass have a shorter interpulse
phase and smaller intershell layers (see \tab{tab:cm-l}), as it is the
case at larger metallicity. For the mass loss law chosen here the
more massive models experience more DUP events during which less
processed material is mixed into the surface. This in addition to the larger
dilution factor of the massive AGB stars due to their larger envelope
mass, leads to the prediction that more massive AGB stars will
generally have smaller overproduction factors, for example for the
overall enrichment in the sum of CNO elements.

In addition to the mixing properties the temperatures at the nuclear
production sites determine the chemical enrichment of the
stars. In \tab{tab:struc-tp} the temperatures in the H-shell, the
He-shell as well as the temperature at the bottom of the envelope
convection and the bottom of the PDCZ are given. The temperature in
the He-shell and the temperature at the bottom of the envelope
convection zone vary substantially and the values given in the table
should be considered typical numbers taken approximately when half of
the interpulse phase has passed. The last line for each sequence gives
the core mass and envelope mass when the computations have been
stopped. The subsequent evolution would lead into the post-AGB and
central star of planetary nebula phase of evolution. The sequences have
been stopped before this numerically difficult transition phase away
from the AGB starts, but after enough envelope mass has been lost for
reliable yield predictions.

The density at the bottom of the
PDCZ ranges from $4.0\times 10^3 
\mem{g/cm^3}$ to  $4.8\times 10^4 \mem{g/cm^3}$ for the 2\msun\
sequence (E82) while slightly lower densities are encountered at
higher core masses ($2\times 10^3$ -- $1.8\times10^4\mem{g/cm^3}$
for the 6\msun\ sequence E86).

\subsection{Chemical evolution and yields}

In the context of Galactic Chemical Evolution different species evolve
differently as a function of metallicity (time) and depending on their
particular location. In order to choose initial abundances for the
stellar evolution models one can either adopt theoretical initial
abundances from published galactic chemical evolution models
\citep{timmes:95}, or the initial composition is tailored to reproduce
observed abundances at [Fe/H]$=-2.3$.  Unfortunately, both approaches
lead to different results for some important species (like N or Na),
while the theoretical models can not be checked in other cases (like
\nezw). Fortunately the yields of many important species of TP-AGB
stars are dominated by primary nucleosynthesis production. Moderate
uncertainties in the initial abundances will have only little effect
on the results. This is certainly true for the CNO elements.
Therefore, the initial composition is set to metallicity scaled solar abundance distribution. The largest difference between the
initial metallicity at $Z=10^{-4}$ and a solar-scaled initial
composition would concern the light elements deuterium and \hedr.
These should be overabundant at very low metallicity compared to the
solar-scaled values due to big bang nucleosynthesis \citep{walker:91}.
In this study the light elements lithium and \hedr\ are not considered.

IMS experience several kinds of dredge-up events that generate
overabundances in their envelopes. At extremely low metallicity the
second dredge-up after the end of core He-burning plays a more
important role than the first dredge-up after the core H-burning. In
fact in this model set only the 2\msun\ sequence has a first dredge-up
that is deeper than the second dredge-up. The most massive cases do
not even have a first dredge-up and Red Giant Branch evolution phase
\citep{girardi:96}.  He-core burning sets in before the model star reaches the
giant branch and temporarily reverses that evolution (\abb{fig:hrd}). 
After that they evolve directly into early AGB stars and the envelope
becomes convectively unstable for the first time. Although the
repeated third dredge-up events are mainly responsible for the total
chemical envelope enrichment with nucleosynthesis products it is
nevertheless important to document the pollution during the evolution
prior to the thermal pulse AGB.

The overabundances just before the first dredge-up event of the
thermal pulse AGB evolution (\abb{fig:o1tp}) reflect mostly abundance
changes that are secondary in nature. The \czw\ abundance is in all
but the 6\msun\ case depleted by mixing the partially CN cycled
envelope. Accordingly the \nvi\ abundance is enhanced. In the 6\msun\
sequence the He-shell is hot and broad enough after the end of He-core
burning that the descending envelope convection (second dredge-up)
engulfs a small amount of primary \czw. This explains also other small
CNO abundance differences between the 6\msun\ and the less massive
cases. The Ne-Al isotopes are effected by secondary p-capture
nucleosynthesis as well. This is more so the case for larger
masses. In particular \nadr\ is produced from p-captures on Ne
isotopes. For this element the production prior to the TP-AGB is
significant compared to the later production during the TP phase if a
solar-scaled \nezw\ initial abundance is assumed. However, this may
not be correct since standard models of massive stars (the yields of
which may dominate the initial abundance distribution of EMP low- and
intermediate mass stars) predict a lower than solar-scaled \nezw\
abundance.

To compare the overall abundance
evolution as a function of mass the envelope mass averaged mass
fractions have been computed for each sequence and each species $i$ using
\begin{equation}
X_\mem{av}^i = \frac{1}{M_\mem{i}-M_\mem{f}}
\int_{M_\mem{f}}^{M_\mem{i}} X^i(m)\, \mem{d}m 
\end{equation}
where $M_\mem{i}$ and $M_\mem{f}$ are the stellar mass at the
beginning and the end of the AGB evolution phase, and X is the surface
mass fraction at a given time corresponding to the stellar mass.
These averaged abundances in the matter returned to the interstellar
medium together with the adopted initial abundance distribution is
given in \tab{tab:oa-abund}. These averaged overabundances should be
useful for comparison with trends in observed abundances of metal poor
globular clusters or extremely metal-poor halo stars. Note that the
squared bracket notation [X/Fe] is very well approximated by $\log
X_\mem{av}/X_\mem{ini}$ because the initial abundance is solar-scaled.
The average overabundances are displayed for all cases in
\abb{fig:oa11}. Yields according to 
\begin{equation}
p_i = -\int_{M_\mem{f}}^{M_\mem{i}} (X_i(m)-X_\mem{ini})\, \mem{d}m  ,
\end{equation}
are given in 
\tab{tab:yields-abund}\footnote{\tab{tab:yields-abund} available online.}. 

\begin{figure}
\plotone{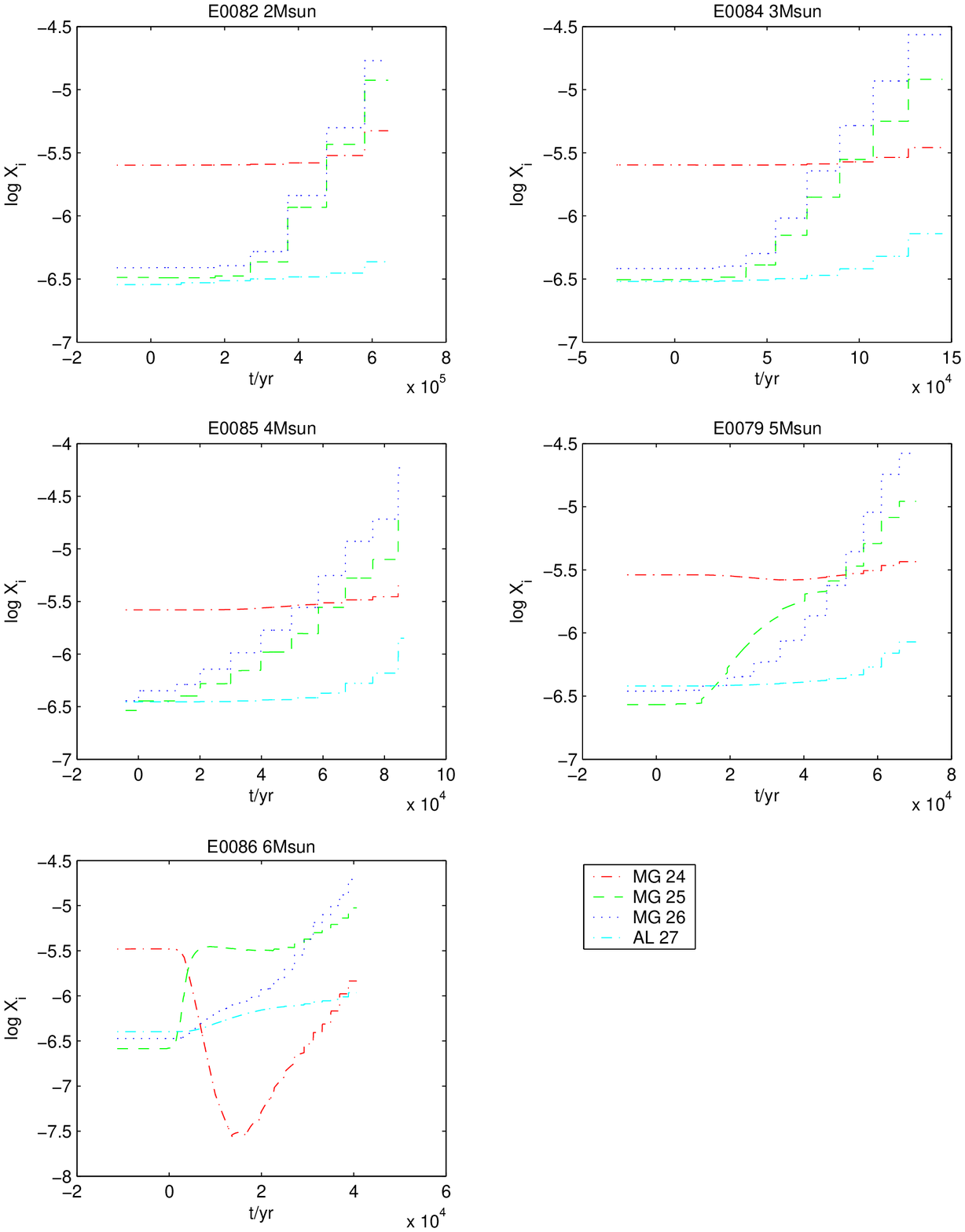} 
\figcaption{ \label{fig:as_multi_MgAl} Abundance evolution of the stable Mg and Al isotopes for all sequences during the TP-AGB evolution.}
\end{figure}

The evolution of surface abundances and stellar parameters for all
tracks is provided in abbreviated online tables (see
\tab{tab:ASPS} for a sample). A graphical representation is given in
\abb{fig:as_multi_CNO} to \ref{fig:as_multi_MgAl}.  For the following
discussion two abbreviations are useful: \emph{LC} refers to the
low-mass cases 2 and 3\msun\ which do not experience hot-bottom
burning, while \emph{HC} refers to the 4-6\msun\ cases in which HBB is
efficient.

\paragraph{Helium} \hevi\ is brought into the envelope by all
dredge-up episodes. In addition HBB in HCs transforms H into
helium. HCs are more efficient \hevi\ producers than LCs, both in
terms of the yields as well as the average abundance in the
ejecta.  The relative importance of the different production
mechanisms is a function of initial mass. HCs generate the vast
majority of \hevi\ during the second dredge-up. Although HBB is most
efficient in the most massive cases the 6\msun\ sequence (E0086), for example,
generates only a few percent of the total \hevi\ overabundance during
the TP-AGB phase. Contrary, the 2\msun\ case (E0082) produces about
two thirds of its \hevi\ overabundance during the TP-AGB by third
dredge-up. The mass dependence of the \hevi\ yields suggests that
for a Salpeter-like intial mass function both HCs and LCs are equally
important producers of \hevi.

\paragraph{Carbon} \czw\ is produced in the He-shell by the
triple-$\alpha$ reaction and dredged-up after the TP.  The LCs show
larger \czw\ overabundances and also larger yields then the HCs.  The
deeper dredge-up (in mass, not in dredge-up parameter $\lambda$) and
the smaller dilution factor due to a smaller envelope mass outweigh
the fact that LCs have less TPs than HCs. \cdr\ is not produced or
destroyed in the 2\msun\ case which does not have HBB. The matter
which is dredged-up from the intershell is void of \cdr\ beacuse of
the large $\cdr(\alpha,\n)\ose$ cross-section.  For this mass the
overabundance is entirely secondary and stems from the pre-TP-AGB
evolution. For larger masses HBB becomes more and more important and
so does the \cdr\ overabundance, which is primary in these cases and
exceeds the enrichment level obtained by the second dredge-up. For $M
\ge 4\msun$ the HBB is efficient so that the \cdr\ overabundance
exceeds the \czw\ overabundance by about a factor of 10.  Both
low-mass stars as well as massive stars produce carbon with much
larger isotopic ratio than this \citep{woosley:02}.  A nuclear
production site like these intermediate mass stars with a small
$\czw/\cdr$ ratio is therefore needed to account for example for the
solar ratio of $\czw/\cdr \simeq 89$.

\paragraph{Nitrogen}
The LCs maintain their isotopic nitrogen enrichment from the first and
second DUP. The nitrogen overabundance in these heavily C-enriched
model stars is therefore of secondary origin. The upper limit of
the nitrogen abundance which can be expected in stars polluted by the
LCs is given by the initial metallicity.

Large primary production of \nvi\ are resulting from combined
efficient DUP and HBB for the HCs only. The 5 and 6\msun\ models even
show a small \nfu\ enrichment. The \nvi/\nfu\ ratio is determined by
the CN cycle equilibrium value. If DUP and HBB are efficient then the
\nvi\ overproduction is so large that some primary \nfu\ is produced
as well. However, this does not address the problem of the nuclear
production site of the solar \nfu\ abundance. In the absence of
detailed models earlier studies have speculated that \nfu\ may be
produced in IMS \citep{timmes:95}. This possibility can be ruled out
at least for  models which employ rather standard physical
assumptions.

Finally note that the carbon to nitrogen ratio in the ejecta of IMS
stars is almost bimodal. The LCs have large [C/N] while the HCs have
small ratios. With standard assumptions on mixing the mass transition
between high and low [C/N] is very sharp. It is therefore not clear
that the simultaneous C and N enrichment observed in many EMP CH-stars
can be explained by mass transfer from companion stars that have
evolved like any of the models presented here.

\paragraph{Oxygen}

\ose\ is enhanced in the envelopes of EMP AGB stars by third
dredge-up. The pattern is the same as in the case \czw: the lower mass
cases bring cumulatively more intershell material into the envelope
compared to the higher mass cases. The origin of \ose\ is the
$\czw(\alpha,\gamma)\ose$ reaction transforming primary \czw\ from the
triple-$\alpha$ reaction. \ose\ is mainly produced in the lower part
of the He-shell where the \hevi\ abundance is declining. During the
He-shell flash the triple-$\alpha$ reaction dominates. The amount of
\ose\ available in the convectively mixed intershell depends on the
treatment of convective boundaries.  With our assumption of
exponential overshoot at the bottom of the PDCZ an \ose\ abundance in
this layer of about $6\%$ by mass is found, depending somewhat on TP
number and mass. As shown in Paper I the prediction of significant
dredge-up of \ose\ in EMP AGB stars is not a result of assuming
overshooting at the bottom of the PDCZ. Even without this kind of
extra-mixing the \ose\ abundance in the PDCZ in EMP stars is of the
order $1-2\%$ by mass. This primary oxygen in the intershell is in any
case several orders of magnitude more abundant than the initial
\ose\ abundance in the envelope.  The importance of \ose\ dredge-up in
AGB stars of the lowest AGB stars is also found in the calculation for
metallicity Z=0 by \citet{siess:02}. For example at the end of the
evolution of their $1.5\msun$ model which does not experience HBB the
surface \czw-mass fraction is $6.9\cdot 10^{-3}$ and that of \ose\ is
$2.4\cdot 10^{-3}$

The \oac\ abundance is depleted rather rapidly for the HCs during HBB
and not replenished because the $\osi(\p,\alpha)\nvi$ reaction exceeds the
$\osi(\p,\alpha)\fac$ reaction by almost two orders of magnitude.

\paragraph{Neon}
A small primary production of \nezwa\ can be noted for all masses
(\abb{fig:as_multi_NeNa}). This is just a curiosity as the nuclear
origin of \nezwa\ is carbon burning in supernova Type II. However, it
is noted that the small increase of \nezwa\ observed in these models
is due to two reactions: $\ose(\alpha,\gamma)\nezwa$ and to a lesser
degree $\ose(\n,\gamma)\osi(\alpha,\n)\nezwa$.

In the LCs the primary \nezw\ can increase the elemental neon
abundance significantly. Both heavy neon isotopes are destroyed by HBB
in the HC cases. During the final pulse cycles HBB becomes less
efficient. Only then is the DUP of \neei\ and \nezw\ more efficient
than the interpulse destruction. In the He-shell flash \neei\ is
mainly made by $\ose(\n,\gamma)\osi(\alpha,\gamma)\neei$ and to a
small extent by $\nezwa(\n,\gamma)\neei$. In the LCs in particular the
\neei\ yields are significant. It is important to note that even if
the $\nezwa(\n,\gamma)\neei$ dominates \neei\ is still primary,
because \nezwa\ is produced in a primary mode, even though this is not
important for the \nezw\ yield at this metallicity.

\paragraph{Sodium}
Sodium is produced by both LCs and HCs by different processes. In HCs
\nezw\ is dredged up from the intershell and then transformed into
\nadr\ during HBB. If, like in the 6\msun\ case, HBB is very efficient
then \nadr\ is first produced and destroyed in the second half of the
interpulse phase. However in all other cases significant amounts of
\nadr\ are produced. In the LCs the main source of \nadr\ is
$\nezw(\n,\gamma)\nadr$.

\paragraph{Magnesium}
The different nuclear production sites during H- and He-shell burning
in TP-AGB stars is discussed in all detail in \citet{karakas:03a}.
Although they have computed models for somewhat higher metallicity all
their basic mechanisms apply equally here. \mgvi\ is only
significantly changed if HBB is very efficient, like in the E0086,
6\msun\ case (\abb{fig:as_multi_MgAl}). Then both \mgfu\ as well as
\mgse\ are produced. This production mode of the neutron heavy
isotopes of magnesium has a distinct signature: the \mgfu\ abundance
is larger than the \mgse\ abundance
\citep{karakas:03a,denissenkov:03a}. Even taking into account
nuclear reaction rate uncertainties does not change this
finding. 

There is a second production site for the neutron
heavy magnesium isotopes. In the He-shell flash convection zone the
two $\alpha$-captures on \nezw\ are almost equally important:
$\nezw(\alpha,\n)\mgfu$ and $\nezw(\alpha,\gamma)\mgse$. However, the
temperature dependence of these two reactions is different. For
$T_8>2.8$ the $(\alpha,\n)$ reaction exceeds the ($\alpha,\gamma$)
reaction, while for lower temperatures \nezw\ is more effectively
transformed into \mgse. This dividing temperature coincides roughly
with the activation temperature of the \nezw\ n-source reaction.
Therefore, for low temperatures in the He-shell flash convection zone
(as can be found in low mass models) one expects only little
neutron-heavy magnesium production favoring \mgse. However, at higher
He-shell flash temperatures should \mgfu\ dominate the heavy Mg
isotope production. 

There is a component missing in this picture, which appears to be
important in EMP stars. \mgfu\ has a very high neutron cross section,
about 600 times larger than that of \czw, about 200 times larger than
\ose, and still about twice as large as the \alse\ n-capture cross
section. The relative number of neutrons actually available for \mgfu\ 
depends not only on the relative value of the cross sections but
equally on the abundance of competing neutron capturing species.
Assuming, for example, that $5\%$ by mass of a $2\%$ mass abundance of
\nezw\ is transformed into \mgfu\ implies a mass abundance for this
isotope of $10\cdot^{-3}$. Thus, for this example $(<\sigma
v>_\mem{(n,\gamma)} Y)_{\mgfu}$ exceeds $(<\sigma v>_\mem{(n,\gamma)}
Y)_{\czw}$. Thus, neutrons are captured by \mgfu\ in significant
numbers if \nezw\ releases neutrons.  In order to check this picture I
have conducted some one-zone nuclear network calculations for typical
conditions found in the He-shell flash convection zone. These
calculations confirm that neutron captures on \mgfu\ are important for
the evolution of the heavy magnesium isotopes in EMP AGB stars. For
temperatures where \nezw\ releases neutrons (in the test calculations
$T_8=3.2$) $X(\mgse)>X(\mgfu)$ if all neutron capture reactions are
considered, but $X(\mgse)<X(\mgfu)$ if the $\mgfu(\n,\gamma)\mgse$
reaction is switched off, as expected from the ratio of two \nezw\ 
$\alpha$ captures. 

It should also be pointed out that this \mgfu\ and \mgse\ from the
He-shell is of primary origin while \mgfu\ and \mgse\ from HBB 
is mainly secondary. There it is produced at the expense of the
initial \mgvi\ abundance, that has its primary origin during carbon and
subsequent neon burning in Type II supernova.

These models show that in fact EMP AGB stars can -- within the
uncertainties of mass loss in particular -- produce primary neutron
heavy magnesium isotopes with a \mgse/\mgfu\ ratio exceeding unity.
This is in particular the case when mass loss is efficient and
prevents HBB from dominating the nucleosynthesis of the neutron-heavy
Mg isotopes. This result should be relevant for the interpretation of
recent isotopic magnesium abundance determinations \citep{yong:03}.

\paragraph{More comparison with other calculations} \citet{ventura:02}
have presented IMS yields from complete stellar evolution calculations
down to a metallicity of Z=0.0002. They focus on the evolution of
helium and the CNO elements and have not included any heavier
elements. Some of their input physics assumptions are different than
those adopted in this study. These include in particular their choices
on the treatment of convection. They use the Full-Spectrum Turbulence
theory \citep{canuto:91} which tends to cause more efficient HBB
compared to the standard MLT. More importantly, they chose to apply no
overshooting or any other means of extra-mixing during the TP-AGB
phase. Even at very low metallicities and the corresponding high core
masses their third dredge-up is as small as $\lambda=0.2 \dots 0.4$.
As a result their average masses in the ejected material of species
like \hevi, \czw\ or \ose\ are significantly smaller than in this
study (\abb{fig:comp}).  A probably small part of this difference may be
attributed to the metallicity difference of a factor of two between
the two sets.  However, while the two sets of models are
quantitatively different the abundance trends with mass are the same.

\section{Conclusion}
\label{sec:concl}
This paper describes the properties of a grid of intermediate mass
stellar evolution models with low metallicity (Z=0.0001).  The
computations include an extensive nuclear reaction network which allow
predictions of yields and surface abundance evolution for a large
number of species. The model also includes neutron capture reactions
which appear to be important for the correct modeling of light neutron
heavy isotopes. A rather small amount of exponential overshooting has
been included. No other non-standard effects (like rotation or
magnetic fields) have been considered. Thus, these are rather standard
models with updated input physics. 

Confronting models and observations at very low metallicity is
especially interesting as Galactic Chemical Evolution had less time to
cover the tracks of individual nuclear production events. In this lies
the importance of studying stars of the lowest metallicity, both
theoretically and observationally. Confronting these standard models
with observations of EMP stars will help to identify what additional
processes are important.

The amount of exponential overshoot in these models is too small to
generate a \cdr\ pocket that could create relevant neutron fluxes.
However, the yields of some neutron-heavy isotopes have been shown in
this paper to be very sensitive to the s-process in EMP stars. In these
models the neutrons come almost exclusively from the He-shell flash
convection zone and \nezw\ as the neutron source. In the future the
role of the \cdr\ pocket for the s-process in general and for the EMP
yields of species like the heavy magnesium isotopes must be studied in
more detail.


\acknowledgments F.\,H.\ appreciates generous support from D.\, A.\ 
VandenBerg through his Operating Grant from the Natural Science and
Engineering Research Council of Canada. This work would have not been
possible without our stellar evolution computer cluster.  I would also
like to thank Maria Lugaro for many important discussions on the
s-process.


\newpage
\begin{deluxetable}{lll} 
\tablecolumns{3} 
\tablecaption{\label{tab:nuc-netw}
Reaction network  \texttt{EDITOR: this table electronically only}} 
\tablehead{ 
\colhead{ID}  &\colhead{Reaction}& \colhead{Ref.} 
} 
\startdata 
\cutinhead{$\p$-capture reactions}
    1 & \verb{2 PROT  ( 0 OOOOO, 0 OOOOO)  1 DEUT { & 1\\
    2 & \verb{1 DEUT  ( 1 PROT , 0 OOOOO)  1 HE  3{ & 1\\
    6 & \verb{1 LI  7 ( 1 PROT , 0 OOOOO)  2 HE  4{ & 1\\
    7 & \verb{1 BE  7 ( 1 PROT , 0 OOOOO)  1 B   8{ & 1\\
   10 & \verb{1 C  12 ( 1 PROT , 0 OOOOO)  1 N  13{ & 1\\
   11 & \verb{1 C  13 ( 1 PROT , 0 OOOOO)  1 N  14{ & 1\\
   12 & \verb{1 N  14 ( 1 PROT , 0 OOOOO)  1 O  15{ & 1\\
   13 & \verb{1 N  15 ( 1 PROT , 1 HE  4)  1 C  12{ & 1\\
   14 & \verb{1 N  15 ( 1 PROT , 0 OOOOO)  1 O  16{ & 1\\
   15 & \verb{1 O  16 ( 1 PROT , 0 OOOOO)  1 F  17{ & 1\\
   16 & \verb{1 O  17 ( 1 PROT , 1 HE  4)  1 N  14{ & 1\\
   17 & \verb{1 O  17 ( 1 PROT , 0 OOOOO)  1 F  18{ & 1\\
   18 & \verb{1 O  18 ( 1 PROT , 1 HE  4)  1 N  15{ & 1\\
   19 & \verb{1 O  18 ( 1 PROT , 0 OOOOO)  1 F  19{ & 1\\
   20 & \verb{1 F  19 ( 1 PROT , 1 HE  4)  1 O  16{ & 1\\
   21 & \verb{1 F  19 ( 1 PROT , 0 OOOOO)  1 NE 20{ & 1\\
   22 & \verb{1 NE 20 ( 1 PROT , 0 OOOOO)  1 NA 21{ & 1\\
   23 & \verb{1 NE 21 ( 1 PROT , 0 OOOOO)  1 NA 22{ & 1\\
   26 & \verb{1 NA 22 ( 1 PROT , 0 OOOOO)  1 MG 23{ & 1\\
   30 & \verb{1 MG 25 ( 1 PROT , 0 OOOOO)  1 AL 26g{& 1\\
   33 & \verb{1 AL 26g( 1 PROT , 0 OOOOO)  1 SI 27{ & 4\\
   35 & \verb{1 AL 27 ( 1 PROT , 1 HE  4)  1 MG 24{ & 1\\
   37 & \verb{1 SI 28 ( 1 PROT , 0 OOOOO)  1 P  29{ & 1\\
   38 & \verb{1 SI 29 ( 1 PROT , 0 OOOOO)  1 P  30{ & 1\\
   39 & \verb{1 SI 30 ( 1 PROT , 0 OOOOO)  1 P  31{ & 1\\
   63 & \verb{1 NE 22 ( 1 PROT , 0 OOOOO)  1 NA 23{ & 1\\
   65 & \verb{1 NA 23 ( 1 PROT , 0 OOOOO)  1 MG 24{ & 1\\
   66 & \verb{1 NA 23 ( 1 PROT , 1 HE  4)  1 NE 20{ & 1\\
   69 & \verb{1 MG 24 ( 1 PROT , 0 OOOOO)  1 AL 25{ & 1\\
   70 & \verb{1 MG 25 ( 1 PROT , 0 OOOOO)  1 AL 26{ & 1\\
   72 & \verb{1 MG 26 ( 1 PROT , 0 OOOOO)  1 AL 27{ & 1\\
   73 & \verb{1 AL 27 ( 1 PROT , 0 OOOOO)  1 SI 28{ & 1\\
   74 & \verb{1 AL 27 ( 1 PROT , 1 HE  4)  1 MG 24{ & 1\\
   79 & \verb{1 B  11 ( 1 PROT , 0 OOOOO)  3 HE  4{ & 1\\
   81 & \verb{1 C  14 ( 1 PROT , 0 OOOOO)  1 N  15{ & a\\
   92 & \verb{1 N  13 ( 1 PROT , 0 OOOOO)  1 O  14{ & 1\\
  101 & \verb{1 P  31 ( 1 PROT , 0 OOOOO)  1 S  32{ & 4\\
\cutinhead{$\alpha$- and \hedr-capture reactions}
    3 & \verb{2 HE  3 ( 0 OOOOO, 2 PROT )  1 HE  4{ & 1\\
    4 & \verb{1 HE  4 ( 1 HE  3, 0 OOOOO)  1 BE  7{ & 1\\
   40 & \verb{3 HE  4 ( 0 OOOOO, 0 OOOOO)  1 C  12{ & 1\\
   41 & \verb{1 C  12 ( 1 HE  4, 0 OOOOO)  1 O  16{ & 1\\
   42 & \verb{1 C  13 ( 1 HE  4, 1 NEUT )  1 O  16{ & 1\\
   43 & \verb{1 N  14 ( 1 HE  4, 0 OOOOO)  1 F  18{ & 1\\
   44 & \verb{1 O  16 ( 1 HE  4, 0 OOOOO)  1 NE 20{ & 1\\
   45 & \verb{1 O  18 ( 1 HE  4, 0 OOOOO)  1 NE 22{ & 1\\
   46 & \verb{1 NE 20 ( 1 HE  4, 0 OOOOO)  1 MG 24{ & 1\\
   48 & \verb{1 NE 21 ( 1 HE  4, 1 NEUT )  1 MG 24{ & 1\\
   49 & \verb{1 NE 22 ( 1 HE  4, 0 OOOOO)  1 MG 26{ & 1\\
   50 & \verb{1 NE 22 ( 1 HE  4, 1 NEUT )  1 MG 25{ & 1\\
   53 & \verb{1 MG 24 ( 1 HE  4, 0 OOOOO)  1 SI 28{ & 3\\
   61 & \verb{1 NE 20 ( 1 HE  4, 1 PROT )  1 NA 23{ & 1\\
   67 & \verb{1 NA 23 ( 1 HE  4, 1 PROT )  1 MG 26{ & 8\\
   68 & \verb{1 MG 24 ( 1 HE  4, 1 PROT )  1 AL 27{ & 1\\
   71 & \verb{1 MG 25 ( 1 HE  4, 1 NEUT )  1 SI 28{ & 1\\
   75 & \verb{1 BE  7 ( 1 HE  4, 0 OOOOO)  1 C  11{ & 1\\
   76 & \verb{1 LI  7 ( 1 HE  4, 0 OOOOO)  1 B  11{ & 1\\
   78 & \verb{1 B  11 ( 1 HE  4, 1 NEUT )  1 N  14{ & 3\\
   80 & \verb{1 C  14 ( 1 HE  4, 0 OOOOO)  1 O  18{ & 9\\
   82 & \verb{1 N  15 ( 1 HE  4, 0 OOOOO)  1 F  19{ & 1\\
   83 & \verb{1 O  17 ( 1 HE  4, 0 OOOOO)  1 NE 21{ & 3\\
   84 & \verb{1 O  17 ( 1 HE  4, 1 NEUT )  1 NE 20{ & 1\\
\cutinhead{$\n$-capture reactions}
   31 & \verb{1 AL 26g( 1 NEUT , 1 PROT )  1 MG 26{ & 3\\
   32 & \verb{1 AL 26g( 1 NEUT , 1 HE  4)  1 NA 23{ & 3\\
   36 & \verb{1 BE  7 ( 1 NEUT , 1 PROT )  1 LI  7{ & 0\\
   47 & \verb{1 NE 20 ( 1 NEUT , 0 OOOOO)  1 NE 21{ & 5\\
   51 & \verb{1 NE 22 ( 1 NEUT , 0 OOOOO)  1 NA 23{ & 6\\
   52 & \verb{1 NA 23 ( 1 NEUT , 0 OOOOO)  1 MG 24{ & 5\\
   54 & \verb{1 MG 24 ( 1 NEUT , 0 OOOOO)  1 MG 25{ & 5\\
   55 & \verb{1 MG 25 ( 1 NEUT , 0 OOOOO)  1 MG 26{ & 5\\
   56 & \verb{1 MG 26 ( 1 NEUT , 0 OOOOO)  1 AL 27{ & 5\\
   57 & \verb{1 AL 27 ( 1 NEUT , 0 OOOOO)  1 SI 28{ & 5\\
   58 & \verb{1 C  12 ( 1 NEUT , 0 OOOOO)  1 C  13{ & 5\\
   62 & \verb{1 NE 21 ( 1 NEUT , 0 OOOOO)  1 NE 22{ & 7\\
   64 & \verb{1 NE 22 ( 1 NEUT , 0 OOOOO)  1 NA 23{ & 6\\
   85 & \verb{1 N  14 ( 1 NEUT , 1 PROT )  1 C  14{ & b\\
   86 & \verb{1 SI 28 ( 1 NEUT , 0 OOOOO)  1 SI 29{ & 5\\
   87 & \verb{1 SI 29 ( 1 NEUT , 0 OOOOO)  1 SI 30{ & 5\\
   88 & \verb{1 SI 30 ( 1 NEUT , 0 OOOOO)  1 P  31{ & 5\\
   89 & \verb{1 N  14 ( 1 NEUT , 0 OOOOO)  1 N  15{ & 5\\
   90 & \verb{1 O  16 ( 1 NEUT , 0 OOOOO)  1 O  17{ & 5\\
   93 & \verb{1 FE 56 ( 1 NEUT , 0 OOOOO)  1 FE 57{ & 5\\
   94 & \verb{1 FE 57 ( 1 NEUT , 0 OOOOO)  1 FE 58{ & 5\\
   95 & \verb{1 FE 58 ( 1 NEUT , 0 OOOOO)  1 CO 59{ & 5\\
   96 & \verb{1 CO 59 ( 1 NEUT , 0 OOOOO)  1 NI 60{ & 5\\
   97 & \verb{1 NI 60 ( 1 NEUT , 0 OOOOO)  1 NI 61{ & 5\\
   98 & \verb{1 NI 61 ( 1 NEUT , 0 OOOOO)  1 NI 62{ & 5\\
  100 & \verb{1 P  31 ( 1 NEUT , 0 OOOOO)  1 S  32{ & 5\\
  106 & \verb{1 NI 58 ( 1 NEUT , 0 OOOOO)  1 NI 59{ & 5\\
  107 & \verb{1 NI 59 ( 1 NEUT , 0 OOOOO)  1 NI 60{ & e\\
  108 & \verb{1 NI 59 ( 1 NEUT , 1 PROT )  1 CO 59{ & 8\\
\cutinhead{$\beta$-decay and $\e^-$-captures}
    5 & \verb{1 BE  7 ( 0 OOOOO, 0 OOOOO)  1 LI  7{ & 1\\
    8 & \verb{1 B   8 ( 0 OOOOO, 0 OOOOO)  2 HE  4{ & 1\\
    9 & \verb{1 B   8 ( 0 OOOOO, 1 PROT )  1 BE  7{ & 1\\
   25 & \verb{1 NA 22 ( 0 OOOOO, 0 OOOOO)  1 NE 22{ & 2\\
   34 & \verb{1 AL 26g( 0 OOOOO, 0 OOOOO)  1 MG 26{ & 2\\
   77 & \verb{1 C  11 ( 0 OOOOO, 0 OOOOO)  1 B  11{ & 2\\
   91 & \verb{1 N  13 ( 0 OOOOO, 0 OOOOO)  1 C  13{ & 2\\
  104 & \verb{1 C  14 ( 0 OOOOO, 0 OOOOO)  1 N  14{ & c\\
  105 & \verb{1 NI 59 ( 0 OOOOO, 0 OOOOO)  1 CO 59{ & d\\
\cutinhead{C-burning}
   59 & \verb{2 C  12 ( 0 OOOOO, 1 PROT )  1 NA 23{ & 3\\
   60 & \verb{2 C  12 ( 0 OOOOO, 1 HE  4)  1 NE 20{ & 3\\
\enddata 
\\
1: NACRE adotped; 2: \citet{horiguchi:96}; 3: \citet{caughlan:88};
4: \citet{iliadis:01};
5: \citet{bao:00};
6: \citet{beer:01};
7: \citet{bao:00a};
8: Hauser-Feshbach \citep{jorissen:01};
9: see \citet{jorissen:01};
a: \citet{wiescher:90};
b: \citet{koehler:89};
c: \citet{takahashi:87};
d: \citet{goriely:99};
e: Hauser-Feshbach \citep{jorissen:01}
\end{deluxetable} 

\begin{deluxetable}{rrrrrrrrrrr} 
\tablecolumns{11} 
\tablecaption{\label{tab:cm-l}
Model properties I} 
\tablehead{ 
\colhead{ID} & \colhead{$M_\mem{ini}$}   &
\colhead{$m_\mem{c}$$^a$}& \colhead{$\log L_\star$$^b$}
& \colhead{$R_\star$$^b$}& \colhead{$N_\mem{TD}$$^c$}&
\colhead{$t_\mem{TP1}$$^d$}&\colhead{$\Delta
M_\mem{Dmax}$$^e$}
&\colhead{$M_\mem{D}$$^f$}&\colhead{$t_\mem{ip}$$^g$}& \colhead{$M_\mem{lost}$$^h$}\\

\colhead{}         & \colhead{$\msun$}   &
\colhead{$m_\mem{c}$}& \colhead{$\lsun$}
& \colhead{$\rsun$}& \colhead{$N_\mem{TD}$}&
\colhead{$10^6\jahre$}&\colhead{$10^{-2}\msun$}
&\colhead{$10^{-2}\msun$}&\colhead{$\jahre$}   & \colhead{$\msun$}}
\startdata 
E82&$2$&$0.626 \pm0.006$&$ 3.96$ &$180 $&$6$ &$799.1$&$1.171$&$5.038$&$82731$&$1.34$\\
E84&$3$&$0.814 \pm0.006$&$ 4.25$ &$250 $&$9$ &$277.1$&$0.507$&$3.465$&$14116$&$2.13$\\
E85&$4$&$0.887 \pm0.002$&$ 4.40$ &$300 $&$10$&$151.0$&$0.367$&$2.707$&$8457$&$3.05$\\
E79&$5$&$0.947 \pm0.001$&$ 4.63$ &$360 $&$12$&$97.2$ &$0.308$&$2.422$&$5281$&$3.99$\\
E86&$6$&$1.033 \pm0.001$&$ 4.70$ &$420 $&$15$&$68.6$ &$0.140$&$1.241$&$2416$&$4.83$\\
\enddata 
\\
$^a$mass of H-free core, range gives total range of core mass
evolution, $^b$approximate average for the entire sequence, $^c$number
of TPs with DUP, $^d$ time at first TP, $^e$ maximum dredged-up mass
after a single TP, $^f$ total dredged-up mass all TPs, $^g$ average
interpulse duration of TPs with DUP, $^h$ total mass lost=M$_\star$ at first TP minus $m_\mem{HTP}$ at end of AGB.
\end{deluxetable} 

\begin{deluxetable}{rrrrrrrrrr} 
\tablecolumns{10} 
\tablecaption{\label{tab:struc-tp} Model properties II} \tablehead{
  \colhead{TP}
  &\colhead{$t_\mem{TP}$$^a$}&\colhead{$T_\mem{FBOT}$$^b$}
  &\colhead{$T_\mem{HES}$$^c$}&\colhead{$T_\mem{HS}$$^d$}&
  \colhead{$T_\mem{CEB}$$^e$}&\colhead{$m_\mem{FBOT}$$^f$}
  &\colhead{$m_\mem{HTP}$$^g$}&\colhead{$m_\mem{DMAX}$$^h$}
  &\colhead{$M_\star$$^i$} \\
} \startdata \cutinhead{E82, M=2\msun}
$ 1$&$    3.97E-02$&$  8.32$&$  8.10 $&$ 7.83$&$  6.31$&$ 0.60550$&$ 0.62559$&$ 0.62740$&$ 1.97251 $\\
$ 2$&$    8.23E+04$&$  8.42$&$  8.09 $&$ 7.81$&$  6.48$&$ 0.60264$&$ 0.62884$&$ 0.62524$&$ 1.96576 $\\
$ 3$&$    1.74E+05$&$  8.44$&$  7.93 $&$ 7.80$&$  6.65$&$ 0.60278$&$ 0.63041$&$ 0.62403$&$ 1.93878 $\\
$ 4$&$    2.70E+05$&$  8.46$&$  7.92 $&$ 7.80$&$  6.74$&$ 0.60424$&$ 0.63132$&$ 0.62279$&$ 1.86456 $\\
$ 5$&$    3.71E+05$&$  8.48$&$  7.93 $&$ 7.79$&$  6.79$&$ 0.60230$&$ 0.63193$&$ 0.62088$&$ 1.70939 $\\
$ 6$&$    4.76E+05$&$  8.49$&$  7.93 $&$ 7.79$&$  6.70$&$ 0.60210$&$ 0.63144$&$ 0.61973$&$ 1.44626 $\\
$ 7$&$    5.79E+05$&$  8.50$&$  7.94 $&$ 7.78$&$  6.20$&$ 0.60065$&$ 0.63111$&$ 0.62021$&$ 1.00773 $\\
\multicolumn{3}{l}{End of AGB}&& & & &$ 0.62768$&$ $&$        0.62823$\\
\cutinhead{E84, M=3\msun}
$  1$&$    2.05E-04$&$  8.38$&$  8.16$&$  7.91 $&$ 6.70$&$ 0.81151$&$ 0.81896$&$ 0.81825 $&$2.94292$\\
$  2$&$    1.13E+04$&$  8.43$&$  8.14$&$  7.90 $&$ 6.99$&$ 0.81156$&$ 0.81945$&$ 0.81711 $&$2.92614$\\
$  3$&$    2.40E+04$&$  8.46$&$  7.98$&$  7.89 $&$ 7.19$&$ 0.81103$&$ 0.81898$&$ 0.81573 $&$2.88628$\\
$  4$&$    3.86E+04$&$  8.48$&$  7.99$&$  7.88 $&$ 7.34$&$ 0.81034$&$ 0.81819$&$ 0.81430 $&$2.81285$\\
$  5$&$    5.46E+04$&$  8.50$&$  8.00$&$  7.88 $&$ 7.47$&$ 0.80858$&$ 0.81726$&$ 0.81274 $&$2.68487$\\
$  6$&$    7.17E+04$&$  8.51$&$  8.01$&$  7.88 $&$ 7.56$&$ 0.80752$&$ 0.81618$&$ 0.81129 $&$2.50025$\\
$  7$&$    8.94E+04$&$  8.52$&$  8.01$&$  7.88 $&$ 7.57$&$ 0.80644$&$ 0.81508$&$ 0.81011 $&$2.23381$\\
$  8$&$    1.08E+05$&$  8.53$&$  8.08$&$  7.88 $&$ 7.41$&$ 0.80549$&$ 0.81414$&$ 0.80907 $&$1.87774$\\
$  9$&$    1.27E+05$&$  8.54$&$  8.08$&$  7.87 $&$ 6.49$&$ 0.80465$&$ 0.81331$&$ 0.80830 $&$1.40821$\\
\multicolumn{3}{l}{End of AGB}&& & & &$ 0.81251$&$ $&$0.81343$\\
\cutinhead{E85, M=4\msun}
$  1$&$    0.00E+00 $&$ 8.47$&$  8.35$&$  7.93 $&$ 7.72$&$ 0.88627$&$ 0.89069$&$ 0.88702 $&$3.93490$\\
$  2$&$    1.19E+04 $&$ 8.48$&$  8.34$&$  7.94 $&$ 7.82$&$ 0.88628$&$ 0.89012$&$ 0.88756 $&$3.81435$\\
$  3$&$    2.01E+04 $&$ 8.50$&$  8.36$&$  7.94 $&$ 7.87$&$ 0.88570$&$ 0.88974$&$ 0.88648 $&$3.65361$\\
$  4$&$    3.00E+04 $&$ 8.50$&$  8.06$&$  7.94 $&$ 7.89$&$ 0.88547$&$ 0.88926$&$ 0.88611 $&$3.36429$\\
$  5$&$    3.98E+04 $&$ 8.52$&$  8.06$&$  7.94 $&$ 7.89$&$ 0.88511$&$ 0.88888$&$ 0.88563 $&$3.00263$\\
$  6$&$    4.98E+04 $&$ 8.53$&$  8.06$&$  7.94 $&$ 7.87$&$ 0.88478$&$ 0.88847$&$ 0.88566 $&$2.55257$\\
$  7$&$    5.85E+04 $&$ 8.54$&$  8.07$&$  7.93 $&$ 7.81$&$ 0.88428$&$ 0.88824$&$ 0.88518 $&$2.12662$\\
$  8$&$    6.73E+04 $&$ 8.55$&$  8.08$&$  7.92 $&$ 7.58$&$ 0.88380$&$ 0.88785$&$ 0.88481 $&$1.72477$\\
$  9$&$    7.62E+04 $&$ 8.55$&$  8.11$&$  7.91 $&$ 6.42$&$ 0.88337$&$ 0.88752$&$ 0.88621 $&$1.36279$\\
$ 10$&$    8.44E+04 $&$ 8.55$&$  8.00$&$  7.90 $&$ 5.23$&$ 0.88469$&$ 0.88843$&$ 0.88747 $&$0.98297$\\
\multicolumn{3}{l}{End of AGB}&& & & &$ 0.88783$&$ $&$0.88916$\\
\cutinhead{E79, M=5\msun}
$  1$&$   -8.10E+01 $&$ 8.36$&$  8.20$&$  7.99 $&$ 7.38$&$ 0.94796$&$ 0.95020$&$ 0.95024 $&$4.93359$\\
$  2$&$    2.64E+03 $&$ 8.40$&$  8.18$&$  7.99 $&$ 7.50$&$ 0.94837$&$ 0.95075$&$ 0.95013 $&$4.91927$\\
$  3$&$    5.32E+03 $&$ 8.44$&$  8.08$&$  7.97 $&$ 7.96$&$ 0.94819$&$ 0.95083$&$ 0.94774 $&$4.90782$\\
$  4$&$    1.22E+04 $&$ 8.48$&$  8.09$&$  7.98 $&$ 7.94$&$ 0.94727$&$ 0.94991$&$ 0.94701 $&$4.79325$\\
$  5$&$    1.92E+04 $&$ 8.50$&$  8.10$&$  7.98 $&$ 7.89$&$ 0.94655$&$ 0.94907$&$ 0.94645 $&$4.56126$\\
$  6$&$    2.64E+04 $&$ 8.51$&$  8.10$&$  7.98 $&$ 7.96$&$ 0.94605$&$ 0.94843$&$ 0.94597 $&$4.13737$\\
$  7$&$    3.35E+04 $&$ 8.52$&$  8.10$&$  7.98 $&$ 7.95$&$ 0.94558$&$ 0.94795$&$ 0.94560 $&$3.54257$\\
$  8$&$    4.02E+04 $&$ 8.52$&$  8.10$&$  7.98 $&$ 7.94$&$ 0.94529$&$ 0.94754$&$ 0.94555 $&$2.90971$\\
$  9$&$    4.62E+04 $&$ 8.53$&$  8.10$&$  7.97 $&$ 7.92$&$ 0.94510$&$ 0.94740$&$ 0.94550 $&$2.38893$\\
$ 10$&$    5.14E+04 $&$ 8.54$&$  8.11$&$  7.96 $&$ 7.86$&$ 0.94494$&$ 0.94730$&$ 0.94547 $&$2.01317$\\
$ 11$&$    5.61E+04 $&$ 8.55$&$  8.11$&$  7.95 $&$ 7.71$&$ 0.94468$&$ 0.94723$&$ 0.94531 $&$1.72970$\\
$ 12$&$    6.10E+04 $&$ 8.56$&$  8.12$&$  7.94 $&$ 7.35$&$ 0.94440$&$ 0.94706$&$ 0.94524 $&$1.49238$\\
$ 13$&$    6.58E+04 $&$ 8.56$&$  8.13$&$  7.94 $&$ 5.96$&$ 0.94431$&$ 0.94691$&$ 0.94614 $&$1.25506$\\
$ 14$&$    7.03E+04 $&$ 8.56$&$  - $&$  -  $&$ - $&$ 0.94486$&$
  0.94756$&$ -  $&$1.01774$\\
\multicolumn{3}{l}{End of AGB}&& & & &$0.94496$&$ $&$0.94613$\\
\cutinhead{E86, M=6\msun}
$  1$&$   -2.29E-01 $&$ 8.37$&$  8.22$&$  8.03 $&$ 7.92$&$ 1.03266$&$ 1.03362$&$ 1.03365 $&$5.86578$\\
$  2$&$    1.05E+03 $&$ 8.39$&$  8.21$&$  8.04 $&$ 7.96$&$ 1.03297$&$ 1.03397$&$ 1.03391 $&$5.85307$\\
$  3$&$    2.11E+03 $&$ 8.41$&$  8.21$&$  8.05 $&$ 7.98$&$ 1.03323$&$ 1.03426$&$ 1.03414 $&$5.83891$\\
$  4$&$    3.17E+03 $&$ 8.43$&$  8.14$&$  8.05 $&$ 8.00$&$ 1.03348$&$ 1.03450$&$ 1.03323 $&$5.82221$\\
$  5$&$    6.54E+03 $&$ 8.45$&$  8.13$&$  8.05 $&$ 8.02$&$ 1.03327$&$ 1.03424$&$ 1.03306 $&$5.59171$\\
$  6$&$    1.00E+04 $&$ 8.49$&$  8.11$&$  8.05 $&$ 8.03$&$ 1.03290$&$ 1.03403$&$ 1.03263 $&$5.16970$\\
$  7$&$    1.36E+04 $&$ 8.50$&$  8.12$&$  8.04 $&$ 8.03$&$ 1.03255$&$ 1.03358$&$ 1.03246 $&$4.45788$\\
$  8$&$    1.70E+04 $&$ 8.53$&$  8.12$&$  8.04 $&$ 8.02$&$ 1.03226$&$ 1.03340$&$ 1.03236 $&$3.74248$\\
$  9$&$    1.99E+04 $&$ 8.53$&$  8.11$&$  8.03 $&$ 8.02$&$ 1.03215$&$ 1.03328$&$ 1.03226 $&$3.15401$\\
$ 10$&$    2.25E+04 $&$ 8.54$&$  8.12$&$  8.02 $&$ 8.00$&$ 1.03201$&$ 1.03319$&$ 1.03229 $&$2.64631$\\
$ 11$&$    2.50E+04 $&$ 8.54$&$  8.12$&$  8.02 $&$ 7.97$&$ 1.03201$&$ 1.03311$&$ 1.03221 $&$2.33477$\\
$ 12$&$    2.72E+04 $&$ 8.55$&$  8.12$&$  8.01 $&$ 7.94$&$ 1.03192$&$ 1.03312$&$ 1.03221 $&$2.09245$\\
$ 13$&$    2.92E+04 $&$ 8.55$&$  8.12$&$  8.00 $&$ 7.90$&$ 1.03189$&$ 1.03309$&$ 1.03222 $&$1.88476$\\
$ 14$&$    3.12E+04 $&$ 8.56$&$  8.13$&$  7.99 $&$ 7.79$&$ 1.03186$&$ 1.03311$&$ 1.03227 $&$1.71168$\\
$ 15$&$    3.32E+04 $&$ 8.57$&$  8.16$&$  7.98 $&$ 7.63$&$ 1.03180$&$ 1.03312$&$ 1.03281 $&$1.56168$\\
$ 16$&$    3.50E+04 $&$ 8.57$&$  8.16$&$  7.98 $&$ 7.29$&$ 1.03222$&$ 1.03354$&$ 1.03327 $&$1.45783$\\
$ 17$&$    3.69E+04 $&$ 8.57$&$  - $&$  -  $&$ 6.20$&$ 1.03272$&$ 1.03402$&$ 1.03377 $&$1.29629$\\
\multicolumn{3}{l}{End of AGB}&& & & &$1.034962 $&$ $&$1.065924$\\
\enddata
\\
$^a$ time since first TP, $t=0\jahre$ at about the time of the
  He-luminosity maximum during the first TP; $^b$ Largest temperature
  at the bottom of the flash-convection zone; $^c$ Temperature in the
  He-burning shell during the interpulse phase following the thermal
  pulse; $^d$ Temperature in the H-burning shell during the interpulse
  phase following the TP; $^e$  Temperature at the convective envelope
  bottom during the interpulse phase following the TP; $^f$ Mass
  coordinate of the flash-driven convection zone bottom;  $^g$ Mass
  coordinate of the H-free core at the time of the TP; $^h$ Mass
  coordinate of the convective envelope bottom at the end of the DUP
  phase; $^i$ Stellar mass at the TP.
\\
\end{deluxetable} 

\begin{deluxetable}{rlrrrrrr} 
\tablecolumns{8} 
\tablecaption{\label{tab:oa-abund}
Average mass fractions in material returned to the ISM} 
\tablehead{ 
\colhead{No}  &\colhead{species}&\colhead{$X_\mem{ini}$}
&\colhead{E0082}  &\colhead{E0084} &\colhead{E0085}
&\colhead{E0079} &\colhead{ E0086} \\
\colhead{}  &\colhead{}&\colhead{solar-scaled}
&\colhead{2\msun}  &\colhead{3\msun} &\colhead{4\msun}
&\colhead{5\msun} &\colhead{6\msun} 
} 
\startdata 
 2&\verb, PROT  ,&   7.6966E-01 & 7.260E-01 &   7.336E-01 &   6.890E-01 &
6.600E-01 &   6.431E-01 \\
 5&\verb, HE  4 ,&   2.3025E-01 & 2.569E-01 &   2.588E-01 &   3.070E-01 &
3.375E-01 &   3.558E-01 \\
 9&\verb, C  12 ,&   1.4750E-05 & 1.369E-02 &   6.090E-03 &   8.980E-04 &
4.587E-04 &   2.159E-04 \\
10&\verb, C  13 ,&   1.7750E-07 & 5.379E-07 &   6.206E-06 &   8.565E-05 &
4.372E-05 &   3.025E-05 \\
11&\verb, N  14 ,&   4.5551E-06 & 2.065E-05 &   1.742E-05 &   2.331E-03 &
1.680E-03 &   8.002E-04 \\
12&\verb, N  15 ,&   1.7950E-08 & 6.114E-09 &   4.338E-09 &   6.465E-08 &
4.431E-08 &   2.023E-08 \\
13&\verb, O  16 ,&   4.1401E-05 & 2.858E-03 &   1.239E-03 &   6.110E-04 &
3.008E-04 &   1.006E-04 \\
14&\verb, O  17 ,&   1.6750E-08 & 7.067E-07 &   5.453E-07 &   9.978E-07 &
8.188E-07 &   4.392E-07 \\
15&\verb, O  18 ,&   9.3502E-08 & 4.984E-08 &   4.930E-08 &   1.877E-09 &
4.783E-10 &   6.399E-11 \\
17&\verb, NE 20 ,&   7.6501E-06 & 1.548E-05 &   1.837E-05 &   1.454E-05 &
1.168E-05 &   9.676E-06 \\
18&\verb, NE 21 ,&   1.9500E-08 & 1.002E-06 &   9.041E-07 &   2.460E-07 &
1.020E-07 &   3.139E-08 \\
19&\verb, NE 22 ,&   6.1501E-07 & 2.733E-04 &   5.500E-05 &   1.272E-05 &
4.035E-06 &   6.782E-07 \\
21&\verb, NA 23 ,&   1.6550E-07 & 3.004E-06 &   1.737E-06 &   8.356E-06 &
3.645E-06 &   4.964E-07 \\
22&\verb, MG 24 ,&   2.5250E-06 & 3.351E-06 &   2.914E-06 &   3.016E-06 &
2.873E-06 &   2.884E-07 \\
23&\verb, MG 25 ,&   3.3251E-07 & 4.961E-06 &   5.390E-06 &   3.291E-06 &
2.825E-06 &   3.794E-06 \\
24&\verb, MG 26 ,&   3.8101E-07 & 6.937E-06 &   1.155E-05 &   7.589E-06 &
4.351E-06 &   2.657E-06 \\
25&\verb, AL26G ,&   1.3000E-26 & 4.531E-10 &   6.980E-10 &   1.109E-08 &
5.930E-08 &   3.028E-07 \\
26&\verb, AL 27 ,&   2.8550E-07 & 3.639E-07 &   4.798E-07 &   4.679E-07 &
4.608E-07 &   6.706E-07 \\
27&\verb, SI 28 ,&   3.2051E-06 & 3.241E-06 &   3.323E-06 &   3.343E-06 &
3.293E-06 &   3.423E-06 \\
28&\verb, SI 29 ,&   1.6800E-07 & 1.842E-07 &   1.955E-07 &   2.035E-07 &
1.881E-07 &   1.780E-07 \\
29&\verb, SI 30 ,&   1.1550E-07 & 1.244E-07 &   1.353E-07 &   1.442E-07 &
1.320E-07 &   1.240E-07 \\
47&\verb, G  63 ,&   1.7000E-08 & 1.499E-07 &   8.597E-08 &   5.737E-08 &
4.277E-08 &   3.272E-08 \\
48&\verb, L   1 ,&   1.0000E-50 & 1.496E-08 &   1.799E-08 &   1.207E-08 &
6.800E-09 &   3.551E-09 \\
\enddata 
\\
$^a$
\end{deluxetable}

\begin{deluxetable}{rlrrrrrr} 
\tablecolumns{7} 
\tablecaption{\label{tab:yields-abund}
Yields \texttt{EDITOR: this table only in the electronic version}}
\tablehead{ 
\colhead{No}  &\colhead{species}
&\colhead{E0082}  &\colhead{E0084} &\colhead{E0085}
&\colhead{E0079} &\colhead{ E0086} \\
\colhead{}  &\colhead{}
&\colhead{2\msun} &\colhead{3\msun} &\colhead{4\msun}
&\colhead{5\msun} &\colhead{6\msun} 
} 
\startdata 
 2&\verb, PROT  ,& -5.955E-02 &  -7.794E-02 &  -2.493E-01 &  -4.335E-01 &  -6.156E-01\\
 5&\verb, HE  4 ,&  3.633E-02 &   6.174E-02 &   2.372E-01 &   4.238E-01 &   6.104E-01\\
 9&\verb, C  12 ,&  1.867E-02 &   1.314E-02 &   2.731E-03 &   1.754E-03 &   9.783E-04\\
10&\verb, C  13 ,&  4.921E-07 &   1.304E-05 &   2.643E-04 &   1.720E-04 &   1.462E-04\\
11&\verb, N  14 ,&  2.197E-05 &   2.784E-05 &   7.193E-03 &   6.619E-03 &   3.869E-03\\
12&\verb, N  15 ,& -1.616E-08 &  -2.945E-08 &   1.444E-07 &   1.042E-07 &   1.111E-08\\
13&\verb, O  16 ,&  3.846E-03 &   2.590E-03 &   1.761E-03 &   1.025E-03 &   2.879E-04\\
14&\verb, O  17 ,&  9.421E-07 &   1.143E-06 &   3.034E-06 &   3.169E-06 &   2.054E-06\\
15&\verb, O  18 ,& -5.962E-08 &  -9.562E-08 &  -2.833E-07 &  -3.676E-07 &  -4.544E-07\\
17&\verb, NE 20 ,&  1.069E-05 &   2.318E-05 &   2.131E-05 &   1.592E-05 &   9.851E-06\\
18&\verb, NE 21 ,&  1.342E-06 &   1.914E-06 &   7.005E-07 &   3.259E-07 &   5.785E-08\\
19&\verb, NE 22 ,&  3.724E-04 &   1.177E-04 &   3.744E-05 &   1.351E-05 &   3.071E-07\\
21&\verb, NA 23 ,&  3.875E-06 &   3.400E-06 &   2.532E-05 &   1.375E-05 &   1.609E-06\\
22&\verb, MG 24 ,&  1.127E-06 &   8.415E-07 &   1.519E-06 &   1.375E-06 &  -1.088E-05\\
23&\verb, MG 25 ,&  6.320E-06 &   1.094E-05 &   9.149E-06 &   9.848E-06 &   1.683E-05\\
24&\verb, MG 26 ,&  8.952E-06 &   2.417E-05 &   2.229E-05 &   1.569E-05 &   1.107E-05\\
25&\verb, AL26G ,&  6.187E-10 &   1.510E-09 &   3.430E-08 &   2.343E-07 &   1.473E-06\\
26&\verb, AL 27 ,&  1.070E-07 &   4.202E-07 &   5.640E-07 &   6.925E-07 &   1.873E-06\\
27&\verb, SI 28 ,&  4.875E-08 &   2.542E-07 &   4.257E-07 &   3.458E-07 &   1.058E-06\\
28&\verb, SI 29 ,&  2.213E-08 &   5.959E-08 &   1.098E-07 &   7.953E-08 &   4.866E-08\\
29&\verb, SI 30 ,&  1.211E-08 &   4.288E-08 &   8.859E-08 &   6.501E-08 &   4.115E-08\\
47&\verb, G  63 ,&  1.815E-07 &   1.492E-07 &   1.248E-07 &   1.018E-07 &   7.646E-08\\
48&\verb, L   1 ,&  2.043E-08 &   3.891E-08 &   3.732E-08 &   2.687E-08 &   1.727E-08\\
\enddata 
\\
$^a$
\end{deluxetable} 

\begin{deluxetable}{lrrrrrrrr} 
  \tablecolumns{8} \tablecaption{\label{tab:ASPS} Structure and
    abundance evolution of computed tracks$^a$} \tablehead{ \colhead{model}
    & \colhead{$M_\star$}& \colhead{$t/\jahre$}&\colhead{$\log
      \teff$}&\colhead{$\log L$ }&\colhead{H}&\colhead{\hevi} &
    \colhead{\czw}&\colhead{\dots} } \startdata
  $    5029$&$   2.00 $&$-7.18 \cdot 10^5$&$  3.67 $&$ 3.21 $&$ 7.56\cdot 10^{-1}$&$  2.44\cdot 10^{-1} $&$ 7.12\cdot 10^{-6}$&\dots\\
  $    5814$&$   1.99 $&$ 3.23 \cdot 10^2$&$  3.66 $&$ 3.48 $&$ 7.55\cdot 10^{-1}$&$  2.454\cdot 10^{-1} $&$ 7.00\cdot 10^{-6}$&\dots\\
\dots&\dots&\dots&\dots&\dots&\dots&\dots&\dots&\dots\\
  \enddata \\
  $^a$ The complete tables ASPS.E82, ASPS.E84, ASPS.E85, ASPS.E79 and
  ASPS.E86 are available electronically. They contain for each
  sequence several structural and abundance quantities at 70 - 150
  times during the TP-AGB evolution. Specifically these tables
  contain: model number, stellar mass in \msun, age ($t=0$ at first
  TP), $\log \teff$, $\log L/\lsun$ and the surface abundance
  evolution of the 23 species given in \tab{tab:oa-abund}. The models
  in each sequence ahve been selected to reconstruct the abundance
  evolution. All details of the stellar parameter evolution (for
  example the exact stellar luminosity evolution after the thermal
  pulses) are only available from the complete evolution sequence,
  which can be obtained from the author.
\end{deluxetable} 

\newpage

\begin{figure}
\plotone{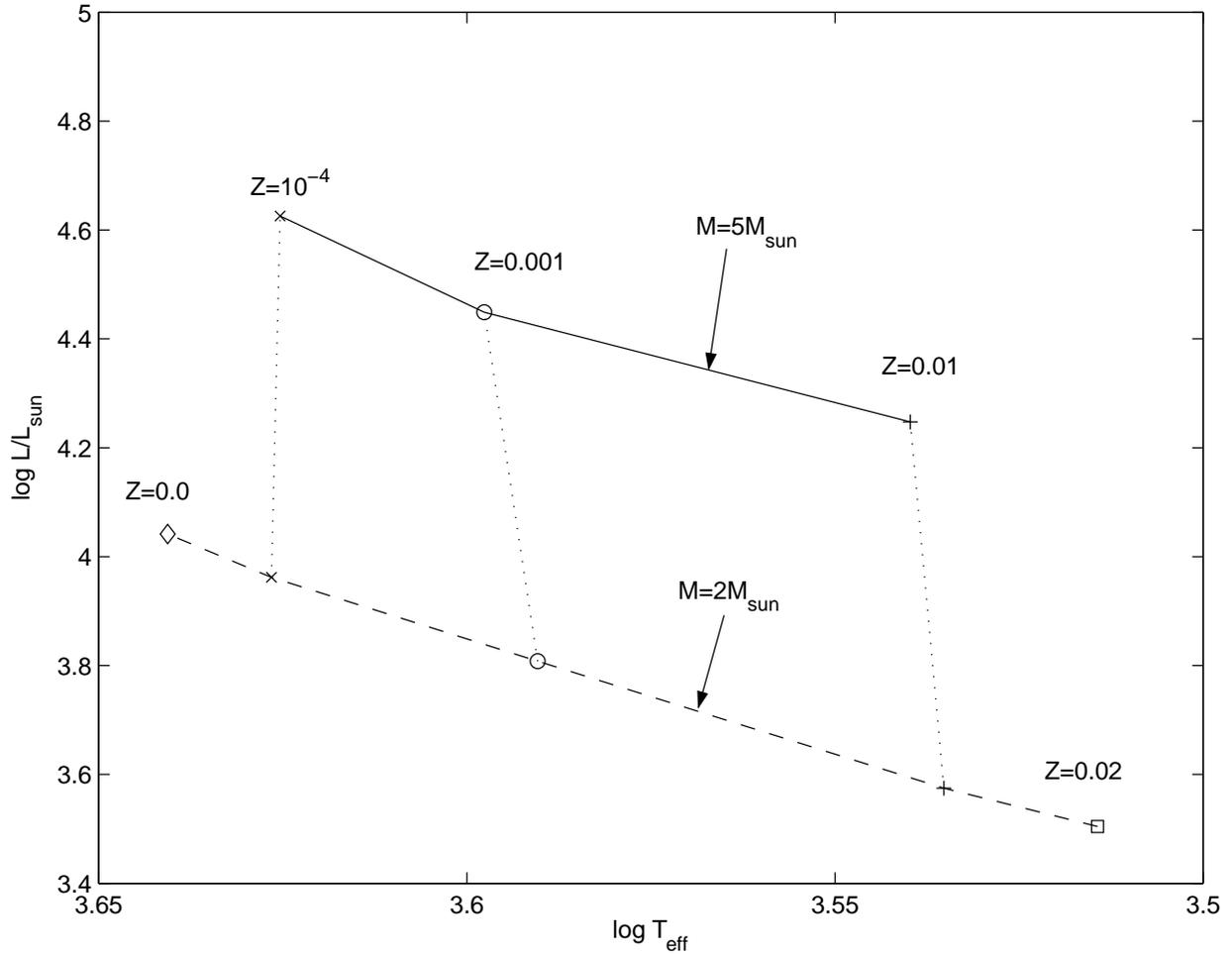} 
\figcaption{ \label{fig:hrd-z} 
Trends of stellar parameters of TP-AGB stellar models for two initial
masses and a range of metallicities. The symbols represent values that
roughly average the variations of the stellar parameters as a function
of the TP cycle as well as the evolution  towards the tip of the AGB.
}\end{figure}
\begin{figure}
\plotone{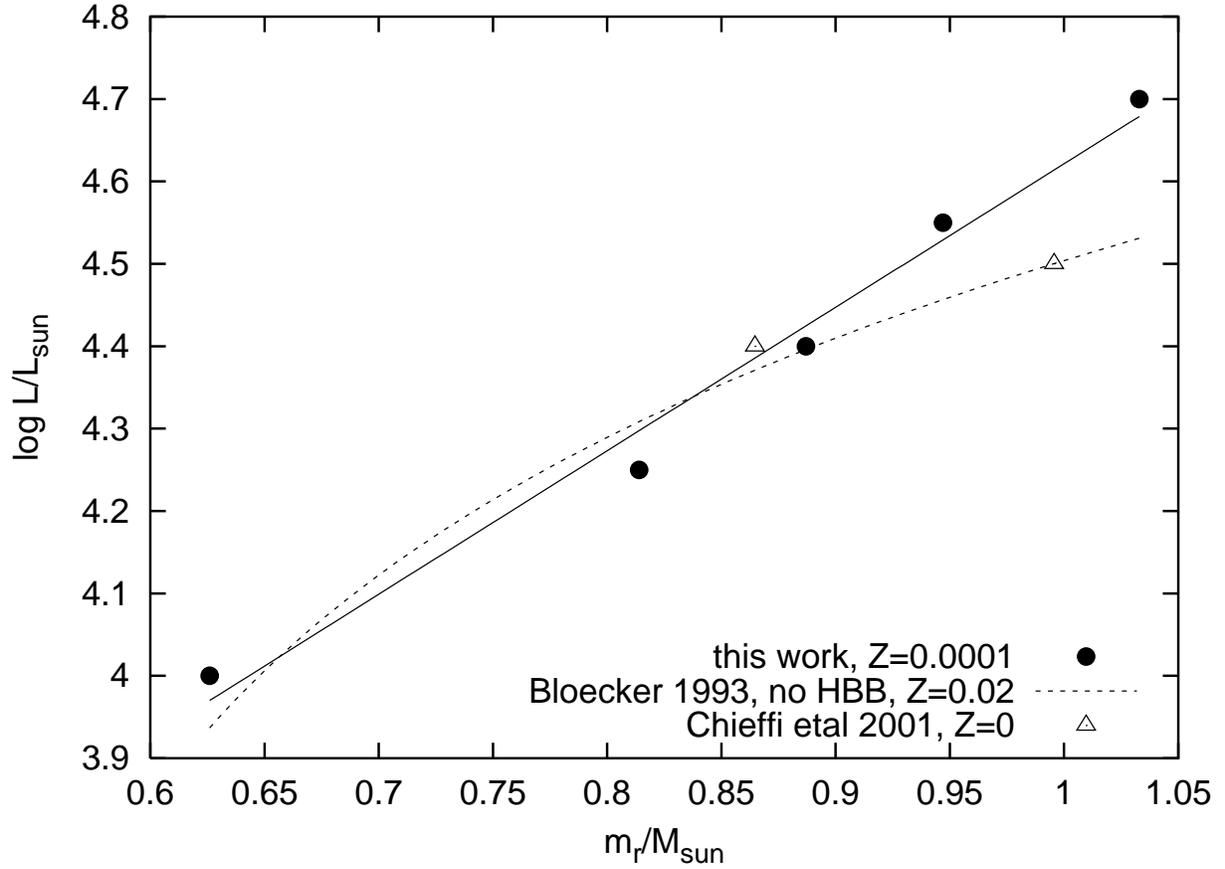} 
\figcaption{ \label{fig:cm-l} 
Average core masses and stellar luminosities of the 5 Z=0.0001
sequences, with a linear $m_\mem{c}-\log L$-fit. As a comparison the
core-mass luminosity relation by  \citet{bloecker:93} for non-HBB
models and two points for Z=0 TP-AGB models during the last computed interpulse
phase by \citet{chieffi:01} are shown.
}\end{figure}

\begin{figure}
\epsscale{0.85}
\plotone{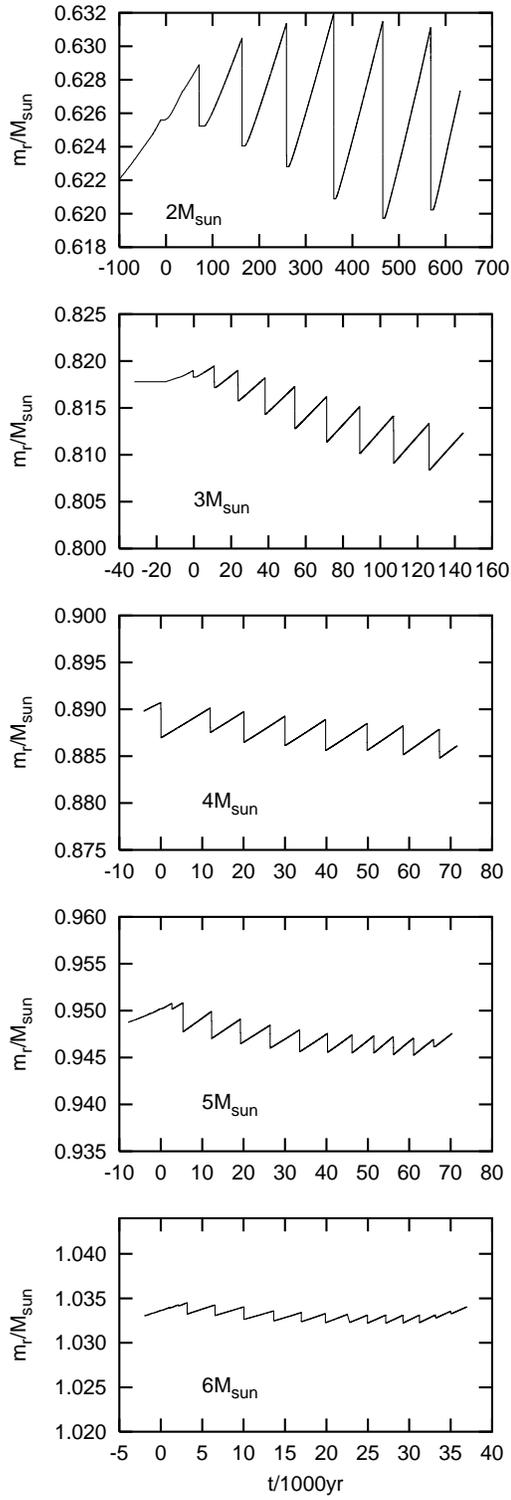} 
\epsscale{1.0}
\figcaption{ \label{fig:t-M} 
Core mass evolution. 
All panels show the same mass range of $\Delta m_\mem{r}=0.025\msun$. 
}\end{figure}

\begin{figure}
\plotone{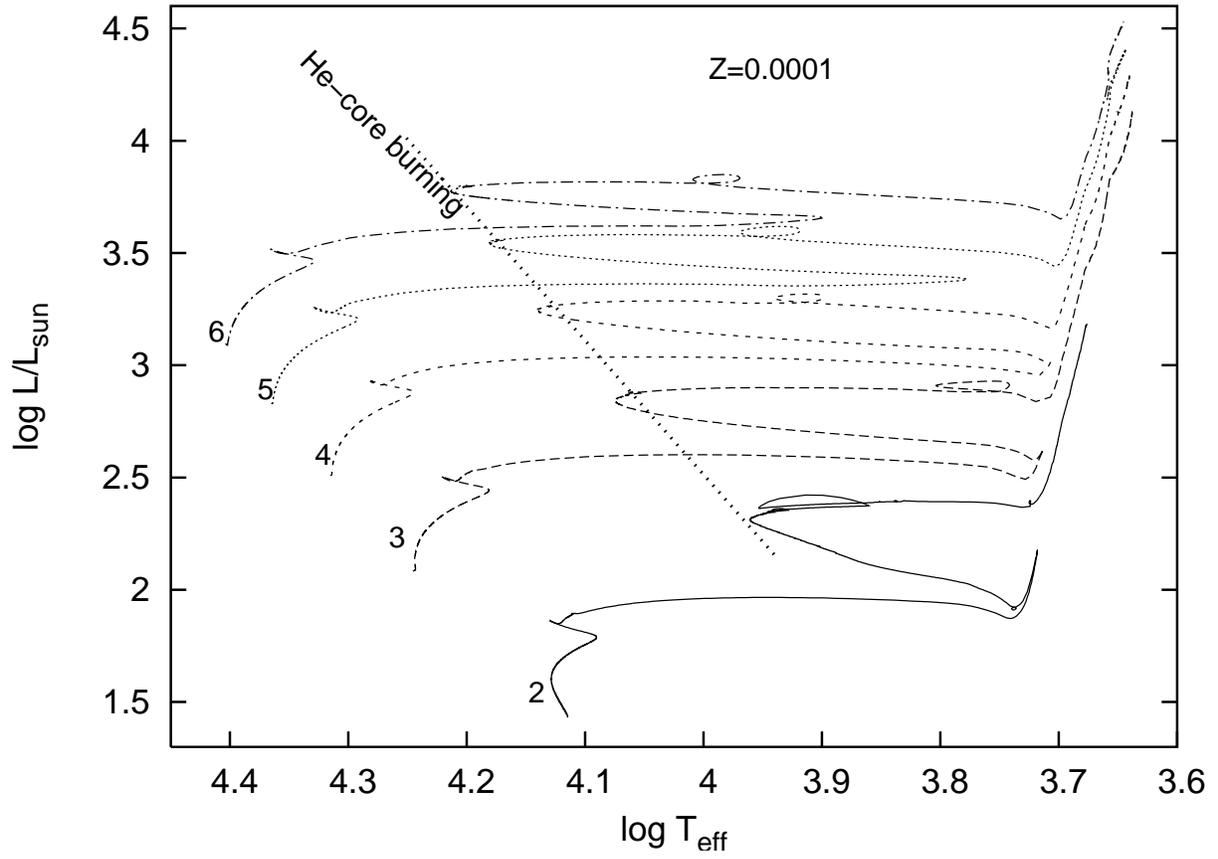} 
\figcaption{ \label{fig:hrd} HRD for the pre-AGB evolution. The
  numeric labels indicate the initial stellar mass of the tracks. Note
  that higher-mass cases do not posses a RGB evolution phase.  }
\end{figure}
\begin{figure}
\plotone{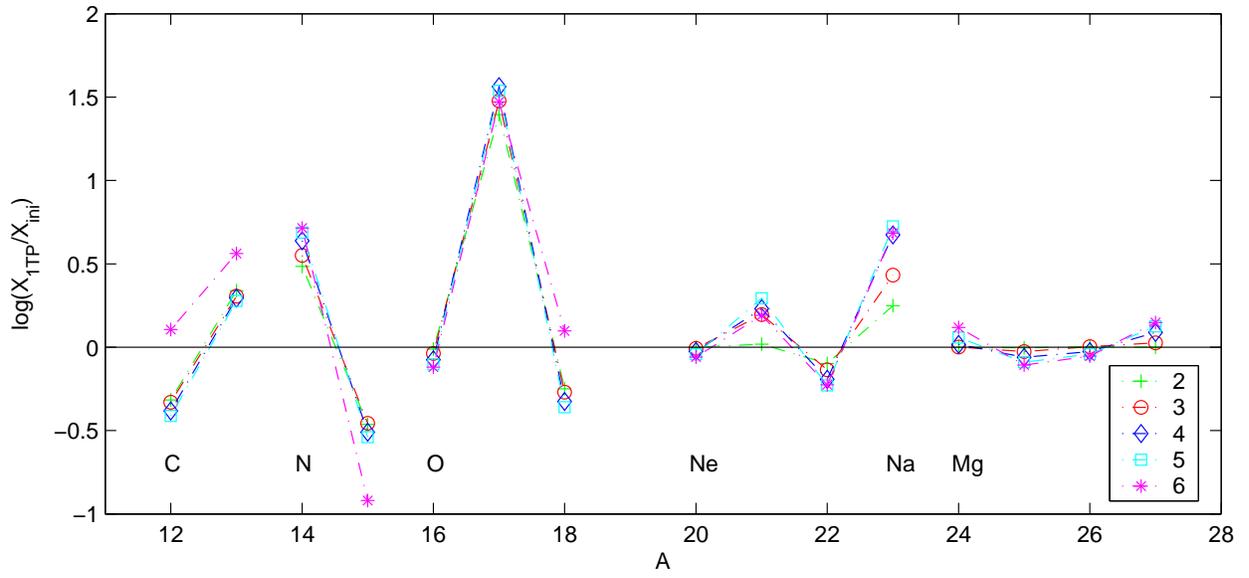} 
\figcaption{ \label{fig:o1tp} Overabundances in the envelope just
  before the first TP, caused by the second dredge-up, and in the
  2\msun\ case through the first DUP. $X_\mem{ini}$ is the
  solar-scaled initial abundance. Lines connect isotopic abundances
  for the respective stellar mass of the elements C, N and O, and for
  the Ne-Na and Mg-Al groups. EDITOR: THIS FIGURE DOUBLE COLUMN
  PLEASE}
\end{figure}

\begin{figure}
\plotone{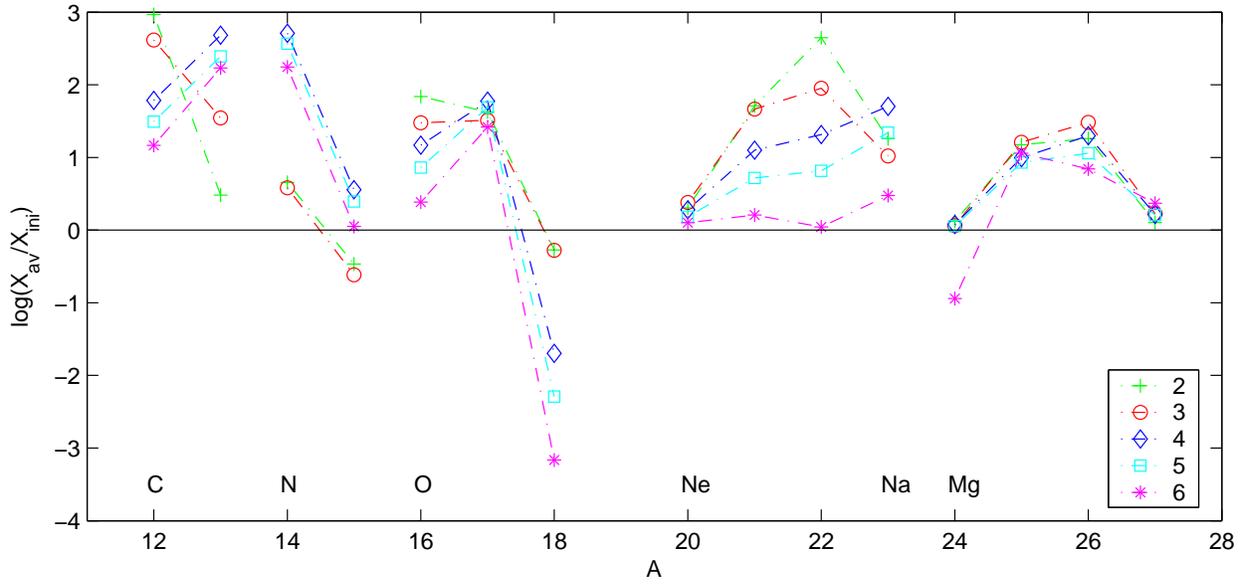} 
\figcaption{ \label{fig:oa11} Overabundances of CNO and Ne-Al isotopes
in material returned to ISM. $X_\mem{ini}$ is solar-scaled and thus
for each species $\log(X_\mem{av}(I)/X_\mem{ini})=\mem{[I/Fe]}$, where
the square brackets have their usual meaning of the logarithmic
abundance ratio relativ to the solar ratio. Lines connect isotopic
abundances for the respective stellar mass of the elements C, N and O,
and for the Ne-Na and Mg-Al groups. EDITOR: THIS FIGURE DOUBLE COLUMN PLEASE}
\end{figure}


\begin{figure}
\plotone{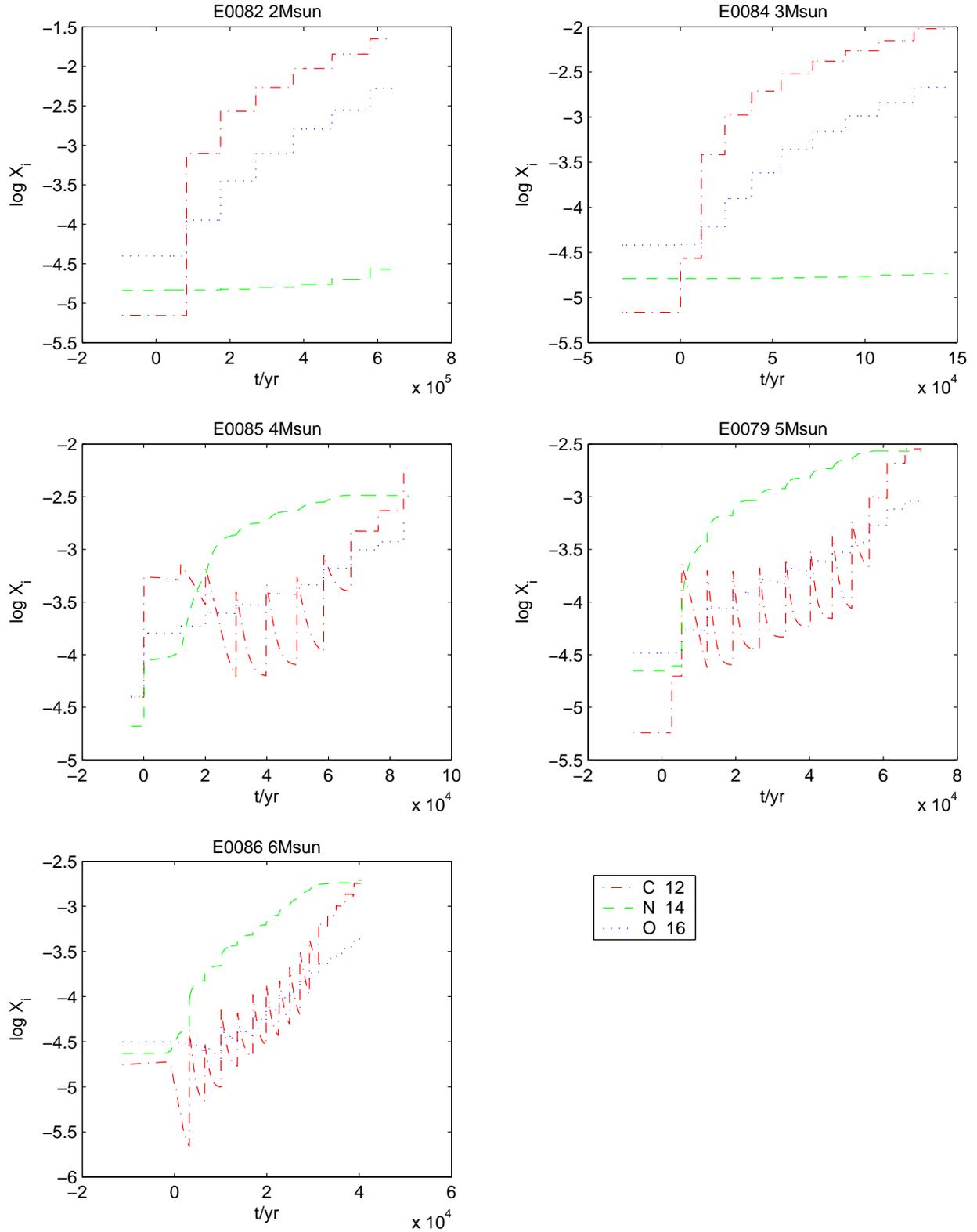} 
\figcaption{ \label{fig:as_multi_CNO} Abundance evolution of the most abundant
CNO isotopes for all sequences during the TP-AGB.}
\end{figure}

\begin{figure}
\plotone{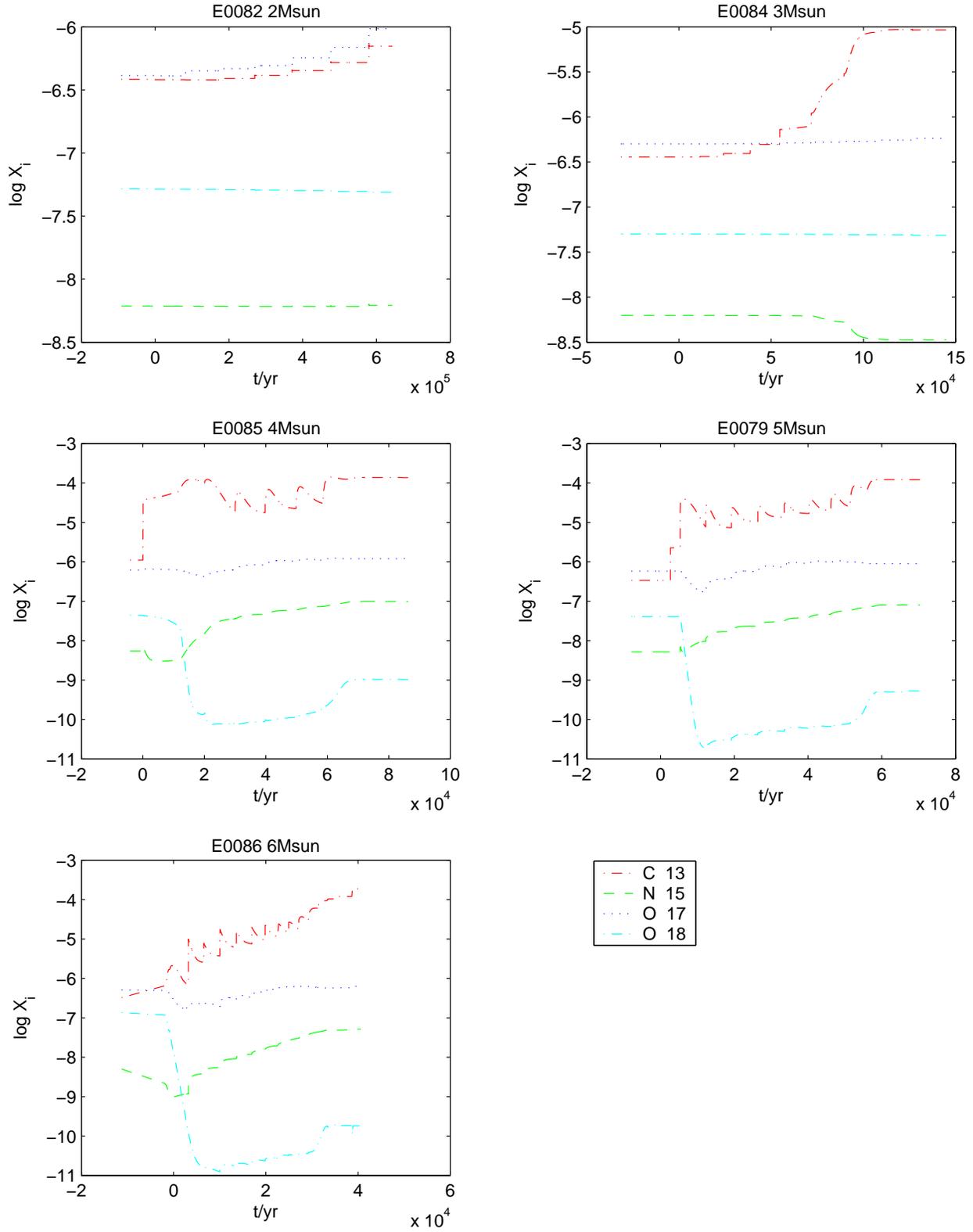} 
\figcaption{ \label{as_multi_CNO_low} Abundance evolution of the less
  abundant CNO isotopes.}
\end{figure}

\begin{figure}
\plotone{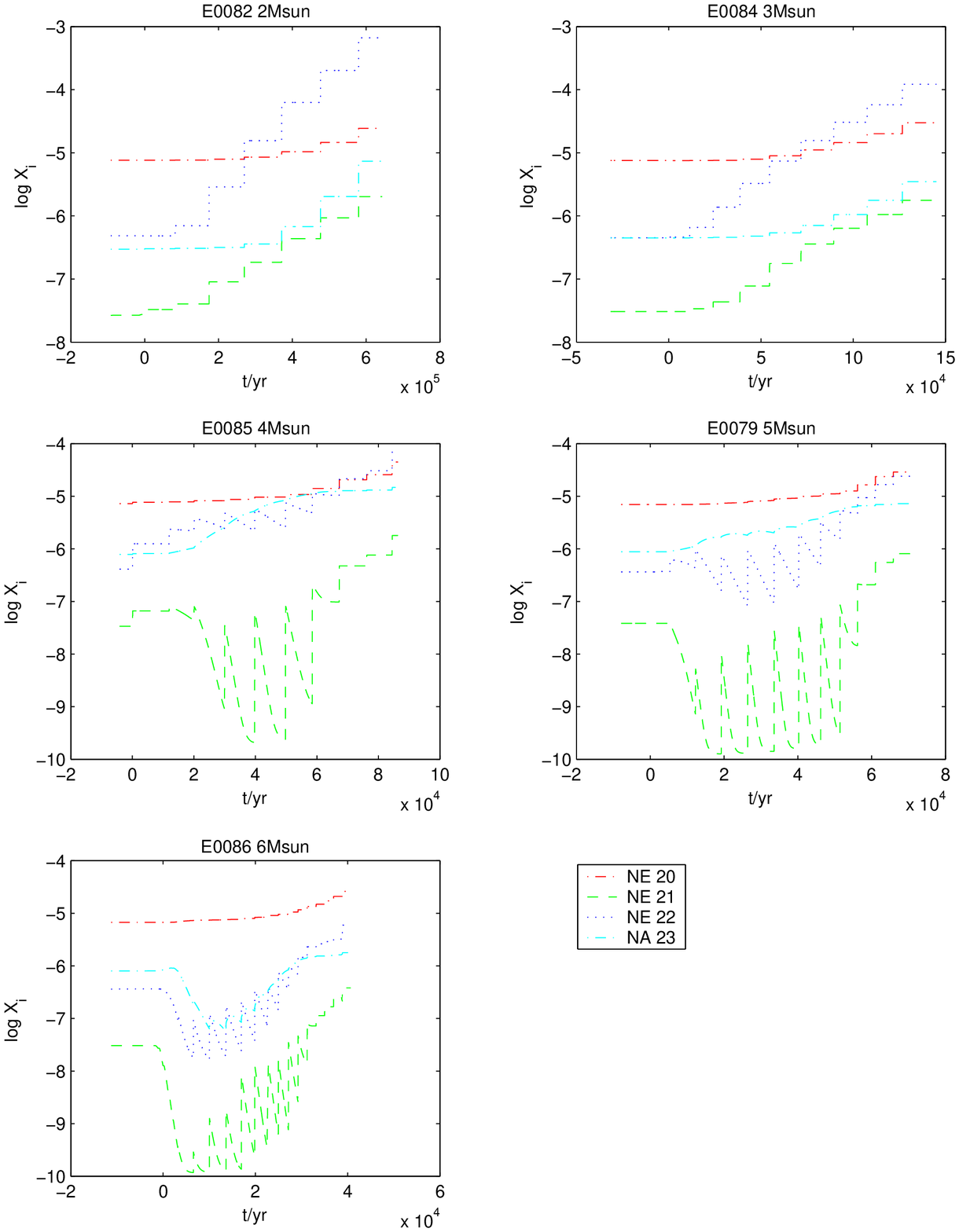} 
\figcaption{ \label{fig:as_multi_NeNa} Abundance evolution of the stable Ne and Na isotopes for all sequences during the TP-AGB evolution.}
\end{figure}

\begin{figure}
\epsscale{0.85}
\plotone{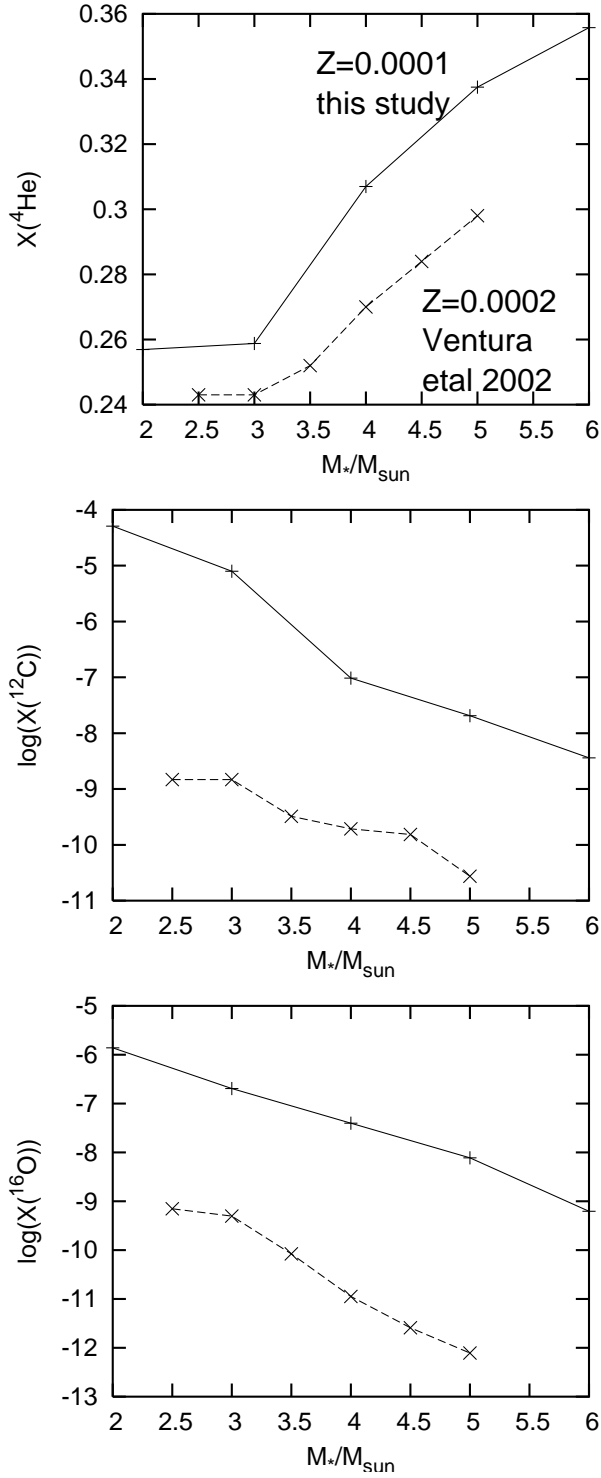} 
\epsscale{1.0}
\figcaption{ \label{fig:comp} Average mass fractions in ejected
  material according to this study compared to
  \citep{ventura:02}. Apart from a small difference in metallicity the
  main difference is in dredge-up efficiency as a result of mixing
  assumptions (see text for details).
}
\end{figure}

\end{document}